\documentclass[3p,preprint]{elsarticle}

\usepackage{amssymb, amsfonts, latexsym, mathtools} %
\usepackage{graphicx, setspace, subfigure}
\usepackage{amsmath,euscript,mathrsfs}
\usepackage[hidelinks]{hyperref} 
\usepackage[usenames,dvipsnames,svgnames]{xcolor}

\hypersetup{
    colorlinks,
    linkcolor={blue!50!black},
    citecolor={blue!50!black},
    urlcolor={blue!50!black}
}

\graphicspath{{Figures/}}

\makeatletter
\def\blfootnote{\gdef\@thefnmark{}\@footnotetext}
\makeatother

\usepackage[all]{hypcap}

\usepackage{array}
\newcolumntype{C}[1]{>{\centering\arraybackslash}m{#1}}

\usepackage[super]{nth}

\usepackage{comment}

\allowdisplaybreaks[1]

\newcommand{\bfu}{\mathbf{u}}
\newcommand{\bfx}{\mathbf{x}}
\newcommand{\Tx}{\tilde{x}}
\newcommand{\Ty}{\tilde{y}}
\newcommand{\Th}{\tilde{h}}
\newcommand{\Tr}{\tilde{r}}
\newcommand{\Tt}{\tilde{t}}

\newcommand{\TU}{\tilde{U}}
\newcommand{\TV}{\tilde{V}}
\newcommand{\sbar}{\bar{s}}
\newcommand{\tbar}{\bar{t}}
\newcommand{\Tbfx}{\tilde{\mathbf{x}}}
\newcommand{\Tbfu}{\tilde{\mathbf{u}}}
\newcommand{\rhat}{\hat{\mathbf{e}}_r}

\newcommand{\Hphi}{\hat{\phi}}

\newcommand{\D}{\mathrm{d}}

\newcommand{\etazero}{\eta^{(0)}}
\newcommand{\etaone}{\eta^{(1)}}
\newcommand{\etatwo}{\eta^{(2)}}

\newcommand{\Te}{\tilde{\eta}}
\newcommand{\He}{\hat{\eta}}
\newcommand{\HU}{\hat{U}}
\newcommand{\Uzero}{U^{(0)}}
\newcommand{\Uone}{U^{(1)}}
\newcommand{\Utwo}{U^{(2)}}

\newcommand{\phizero}{\phi^{(0)}}

\newcommand{\eps}{\varepsilon}
\DeclareMathOperator{\sech}{sech}
\newcommand{\im}{\mathbf{i}}
\newcommand{\F}{\mathcal{F}}

\newcommand{\taumin}{\tau_\mathrm{min}}
\newcommand{\taumax}{\tau_\mathrm{max}}
\newcommand{\ximin}{\xi_\mathrm{min}}
\newcommand{\ximax}{\xi_\mathrm{max}}

\newcommand{\Rmin}{R_\mathrm{min}}
\newcommand{\Rmax}{R_\mathrm{max}}

\newcommand{\Rhat}{\hat{R}}
\newcommand{\That}{\hat{T}}

\usepackage[normalem]{ulem}

\journal{Wave Motion}

\begin{document}

\begin{frontmatter}

\title{Nonlinear concentric water waves of moderate amplitude} %
\author[lbro]{Nerijus Sidorovas}
\author[lbro]{ Dmitri Tseluiko}
\author[njit]{Wooyoung Choi}
\author[lbro]{Karima Khusnutdinova\corref{cor1}}
\ead{K.R.Khusnutdinova@lboro.ac.uk}
\cortext[cor1]{Corresponding author.}
\address[lbro]{Department of Mathematical Sciences, Loughborough University, Loughborough, LE11 3TU, UK}
\address[njit]{Department of Mathematical Sciences, New Jersey Institute of Technology, Newark, NJ 07102-1982, USA}

{\it Dedicated to the memory of Noel Smyth}

\begin{abstract}%
We consider the outward-propagating nonlinear concentric water waves within the scope of the 2D Boussinesq system. The problem is axisymmetric, and we derive the slow radius versions of the cylindrical Korteweg - de Vries (cKdV) and extended cKdV (ecKdV) models. Numerical runs are initially performed using the full axisymmetric Boussinesq system. At some distance away from the origin, we use the numerical solution of the Boussinesq system as the ``initial condition'' for the derived cKdV and ecKdV models. We then compare the evolution of the waves as described by both reduced models and the direct numerical simulations of the axisymmetric Boussinesq system. The main conclusion of the paper is that the extended cKdV model provides a much more accurate description of the waves and extends the range of validity of the weakly-nonlinear modelling to the waves of moderate amplitude.
\end{abstract}


\end{frontmatter}

\section{Introduction}
\label{sec:Intro}

Concentric water waves are a familiar phenomenon to anyone who has ever thrown a stone into water. They are naturally generated in oceans and rivers by some localised disturbances, and they are also important building blocks of more complicated wave patterns such as the Kelvin ship waves (e.g. \cite{W}). Naturally, there have been considerable efforts, both numerical and analytical, to model and study the waves.\\

Theoretical progress was made mainly within the scope of the weakly-nonlinear analysis. The cylindrical (or concentric) Korteweg - de Vries (cKdV) equation
\begin{equation}
2 A_R + 3 A A_{\xi} + \frac 13 A_{\xi \xi \xi} + \frac{A}{R} = 0
\end{equation}
is a universal weakly-nonlinear weakly-dispersive long-wave equation in cylindrical geometry. It was introduced independently by Iordansky (1959) \cite{I} for water waves (potential formulation) and Maxon $\&$ Viecelli (1974) \cite{MV} for waves in plasma. Here $R$ is a slow variable (it can be a slow time or a slow radius), and $\xi$ is, for example, for an outward-propagating wave, a characteristic variable $\xi =r-t$, where $r$ and $t$ are the fast radial and time variables.  Iordansky's model has a similar form, but is written in terms of both characteristic variables of the linear wave equation. Therefore, it is not exactly the same cKdV equation as derived for water waves by Miles (1978) \cite{M} from the Boussinesq system and by Johnson (1980) \cite{J1} from the full Euler equations, but it is an asymptotically equivalent model.  The cKdV equation is an integrable model. Its Lax pair was found by Druma (1976) \cite{D} and Calogero $\&$ Degasperis (1978) \cite{CD}, and interesting exact solutions involving the Airy functions were constructed in the latter paper and by Nakamura $\&$ Chen (1981) \cite{NC}. Considerable efforts were directed towards studying the decay rate of the amplitude of the KdV-type pulses due to cylindrical divergence (see \cite{S, HRS} and references therein). Adiabatic approximations for the slowly varying pulses in the far field (large $R$) were developed in \cite{KK, DPS, J2}, where the cylindrical divergence term has been treated as a perturbation of the KdV equation. \\

The cKdV-type models have also been developed in other physical contexts. In particular, the equation was derived for the concentric internal waves by Lipovskii (1985) \cite{L} and Weidman $\&$ Velarde (1992) \cite{WV}. Important extensions of the models allowing for the presence of a background parallel shear flow were made by Johnson (1990) \cite{J3} for surface waves and Khusnutdinova $\&$ Zhang (2016) \cite{KZ} for internal waves (see also the applications to the modelling of the ring waves in two- and three-layer flows \cite{HKG, TABK}). \\

An extended KdV equation with higher-order nonlinear and dispersive terms was first introduced by Koop and Butler (1981) \cite{KB} for plane internal waves in a two-layer fluid in the situation when the quadratic nonlinearity coefficient is small or vanishing \cite{KB}. This was followed by extensive studies of related models, including the Gardner equation (e.g. \cite{GPTK, OPSS} and references therein). Recently, a similar equation was also derived for the strain waves in an elastic rod by Garbuzov et al (2021) \cite{GBK}. In that context, it was shown that even when the quadratic nonlinearity coefficient was not small,  the extended KdV equation had advantage compared to the KdV equation in the sense that it could accurately describe not only the small amplitude waves, but also the waves of moderate amplitude. Moreover, its soliton solution constructed using the reduction of the extended KdV equation to the Gardner equation by a suitable Kodama-Fokas-Liu near-identity transformation \cite{K, FL} provided a useful initial condition for the modelling of even strongly-nonlinear waves (e.g. table-top solitons). We note that the use of near-identity transformations in the studies of soliton and undular bore solutions of the extended KdV equation was pioneered by Marchant and Smyth, who reduced the equation to the KdV model (e.g. \cite{MS1, MS2}), and that the extended KdV equation has been reently used to describe undular bores in different physical contexts (see, for example, \cite{BS, HRHK}) and references therein.  \\

For concentric waves, a slow-time version of an extended cKdV (ecKdV) equation was recently derived from the potential water waves formulation  by Horikis et al. (2021) \cite{HFMS1} (see also the relevant reviews \cite{KST, HFMS2}). A natural question that can be asked is how accurately the ecKdV equation describes concentric water waves in comparison with the cKdV equation, in particular when the wave
amplitudes are no longer small. This can be effectively answered by comparing the numerical solutions of the cKdV and ecKdV equations with those of their parent systems.  \\

In this paper, we adopt the classical 2D Boussineq equations as a parent model and derive the cKdV and ecKdV equations.
These reduced models are then solved numerically and their solutions are compared with the numerical solutions of the Boussinesq equations to investigate the range of validity of the reduced models. As discussed in Appendix A, different long-wave models could be used, but might be computationally challenging for axisymmetric wave problems.  On the other hand, the Boussinesq equations can be solved accurately using, for example, a pseudospectral method, and provide a good testing ground. \\

\begin{figure}
\centering
\includegraphics[width = 11cm]{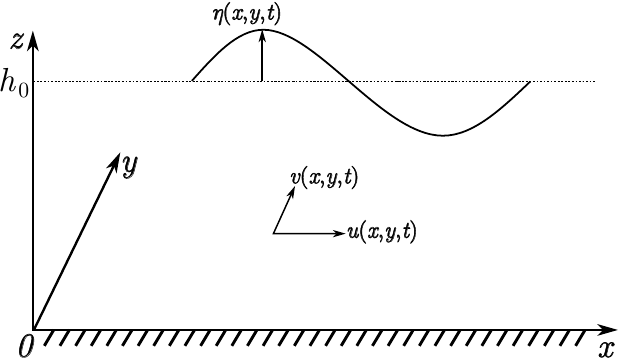}
\caption{The schematic showing 2D surface waves.  The fluid has the unperturbed depth $h_0$ and the free surface elevation $\eta(x,y,t)$. The depth-averaged horizonal speeds in $x$ and $y$ directions  are $u$ and $v$, respectively.}
\label{SurfaceWaveSchematic}
\end{figure}

The rest of the paper is organised as follows. We introduce the axisymmetric Boussinesq system in Section \ref{sec:model} and derive a slow radius version of the extended cKdV equation in Section \ref{sec:cKdV}.  Unlike the leading-order cKdV equation,  the slow time and slow radius versions of the ecKdV equation differ.   The numerical methods used to integrate axisymmetric Boussinesq, cKdV and ecKdV models are described in Section \ref{sec:Num}. In Section \ref{sec:Num1} we compare the predictions of Johnson's asymptotic solution for the cKdV equation \cite{J1} with the results of our direct numerical simulations and clarify the range of validity of this approximation. 
In Section \ref{sec:Num2}, we then consider the same initial value problem and compare the results of direct numerical simulations of the axisymmetric Boussinesq system and the results obtained using both the cKdV and extended cKdV equations. More precisely, in the latter cases 
the reduced models are solved starting at some distance from the origin using the Boussunesq solution at that distance  as an initial condition.
We clarify the range of validity of the models by carefully examining the difference between the results of direct numerical simulations of the parent system and those obtained using the reduced models. Lastly,  we conclude in Section \ref{sec:Conclusion}, and derivation of the slow radius versions of the ecKdV model from the strongly-nonlinear long-wave models (both for outward- and inward-propagating waves) is discussed in the Appendix.   \\



\section{Boussinesq equations}
\label{sec:model}

For our study of concentric water waves, we use the following 2D Boussinesq system (see, for example, \cite{CC1, M1}) 
\begin{equation}
\label{Boussinesq2D}
\begin{cases}
h_t + \nabla \cdot (h \bfu) = 0 , \\
\bfu_t + (\bfu \cdot \nabla)\bfu + g \nabla h = \dfrac{h_0^2}{3} \nabla (\nabla \cdot \bfu_t),
\end{cases}
\end{equation}
where $\bfx = ( x , y )^T$ are the independent horizontal coordinates.
The function $h = h(\bfx,t)$ denotes the local depth of the fluid with $h_0$ being the equilibrium depth, and $\bfu = (u(\bfx,t) , v(\bfx,t) )^T$ being the depth-averaged horizontal velocity vector. The water depth is given by $h = h_0 + \eta$ with $\eta = \eta(\bfx,t)$ denoting the displacement of the surface from the equilibrium (see the schematic in Figure \ref{SurfaceWaveSchematic}). We use this weakly-nonlinear model in cylindrical coordinates as a testing ground and perform the accurate long-time numerical simulations of the propagation of axisymmetric concentric waves using a pseudo-spectral method. Relevant to our current study, perhaps one of the earliest attempts to model concentric water waves was undertaken by Chwang and Wu \cite{CW} in their study of inward-propagating waves within the scope of an axisymmetric Boussinesq system. This study was, however, limited to relatively short time intervals. \\

We non-dimensionalise system (\ref{Boussinesq2D}) using 
\begin{equation}
\eta = \eta_0 \Te, \qquad h = h_0 \Th = h_0 \left( 1 + \dfrac{\eta_0}{h_0} \Te \right), \qquad \bfx = x_0 \Tbfx, \qquad t = t_0\Tt, \qquad \bfu = \dfrac{\eta_0}{h_0} \cdot \dfrac{x_0}{t_0} \Tbfu,
\end{equation}
with $\eta_0$, $h_0$, $x_0$, and $t_0$ being the typical amplitude, depth, length, and time scales, respectively, and also $x_0 / t_0 = \sqrt{gh_0}$ being the linear long-wave speed. System ($\ref{Boussinesq2D}$) takes the following dimensionless form:

\begin{equation}
\label{Boussinesq2Ddimless}
\begin{cases}
\Te_{\Tt} + \tilde{\nabla} \cdot [(1 + \eps \Te) \Tbfu] = 0 , \\[0.3cm]
\Tbfu_{\Tt} + \eps (\Tbfu \cdot \tilde{\nabla}) \Tbfu + \tilde{\nabla} \Te = \dfrac{\delta^2}{3} \tilde{\nabla} (\tilde{\nabla} \cdot \Tbfu_{\Tt}),
\end{cases}
\end{equation}

where $\delta = h_0 / x_0$ and $\eps = \eta_0/h_0$ are the usual long-wavelength and small-amplitude parameters, and the Boussinesq system retains only the leading-order corrections in either of them. In our weakly-nonlinear analysis we will impose the maximal balance condition $\delta^2 = \eps$ in order to retain both the nonlinear and dispersive effects. \\

\begin{figure}
\centering
\includegraphics[width = 5cm]{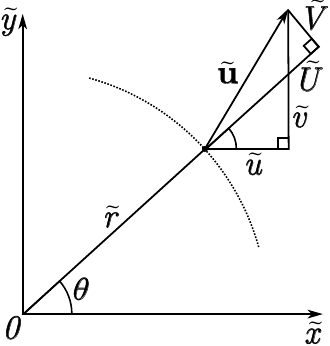}
\caption{Conversion from Cartesian coordinates $\tilde x$, $\tilde y$ (with speeds $\tilde u$, $\tilde v$) to Cylindrical coordinates $\tilde r$, $\theta$ (with radial and tangential speeds as $\tilde U$, $\tilde V$, respectively). }
\label{cylindrical_coords}
\end{figure}

We consider axisymmetric ring waves in the cylindrical coordinate system, and therefore we change variables in $(\ref{Boussinesq2Ddimless})$  as follows (see Figure \ref{cylindrical_coords}):
\begin{equation}
\Tbfx = \begin{pmatrix} \Tx \\ \Ty \end{pmatrix} = \begin{pmatrix} \Tr \cos \theta \\ \Tr \sin \theta \end{pmatrix}, \qquad 
\Tbfu =
\begin{pmatrix} \TU \cos \theta - \TV \sin \theta \\ \TU \sin \theta + \TV \cos \theta \end{pmatrix}.
\end{equation}
Here, $\Tr = \sqrt{\Tx^2 + \Ty^2}>0$ is the distance from the origin, $\theta \in [0,2\pi)$ is the polar angle, and $\TU$ and $\TV$ are the radial and transverse projections of the velocity vector, respectively. Note that $\TU$, $\TV$ are functions of $\Tr$, $\theta$, $\Tt$. For axisymmetric waves, we let $\TV=0$ 
 and omit dependence of $\TU$ on the angle $\theta$, which implies 
\begin{equation}
\Te = \Te(\Tr,\Tt), \qquad \TU = \TU(\Tr,\Tt), \qquad \Tbfu = \TU \rhat ,
\end{equation}
where $\rhat$ is the unit radial vector.
Thus, we reduce system (\ref{Boussinesq2Ddimless}) to the form of the axisymmetric Boussinesq system \cite{CW}:

\begin{equation}
\label{Boussinesq2Dcylindrical}
\begin{cases}
\Te_{\Tt} + \dfrac{1}{\Tr} \bigg[ \Tr (1 + \eps \Te) \TU \bigg]_{\Tr} = 0 , \\[0.5cm]
\TU_{\Tt} + \eps \TU \TU_{\Tr} + \Te_{\Tr} - \dfrac{\eps}{3} \left[ \TU_{\Tt \Tr \Tr} + \dfrac{1}{\Tr} \TU_{\Tt \Tr} - \dfrac{1}{\Tr^2} \TU_{\Tt} \right] = 0.
\end{cases}
\end{equation}

For the long-time simulations of an outward propagating wave it is desirable to switch to a moving reference frame by making the change of variables $(\xi,\tau) = (\Tr - \Tt,\Tt)$ so that the observer moves outward in the radial direction with the linear long-wave speed. This brings system (\ref{Boussinesq2Dcylindrical}) to the following form useful for numerical simulations:

\begin{equation}
\label{Boussinesq2DcylindricalMoving}
\begin{cases}
\Te_{\tau} - \Te_\xi + \dfrac{1}{\xi + \tau} \bigg[ (\xi + \tau) (1 + \eps \Te) \TU \bigg]_{\xi} = 0 , \\[0.5cm]
\TU_{\tau} - \TU_\xi + \eps \TU \TU_{\xi} + \Te_{\xi} - \dfrac{\eps}{3} \left[ \TU_{\tau \xi \xi} - \TU_{\xi \xi \xi} + \dfrac{\TU_{\tau \xi} - \TU_{\xi\xi}}{\xi + \tau} - \dfrac{\TU_{\tau} - \TU_\xi }{(\xi + \tau)^2}\right] = 0.
\end{cases}
\end{equation}

\begin{figure}
\centering
\includegraphics[width = 8cm]{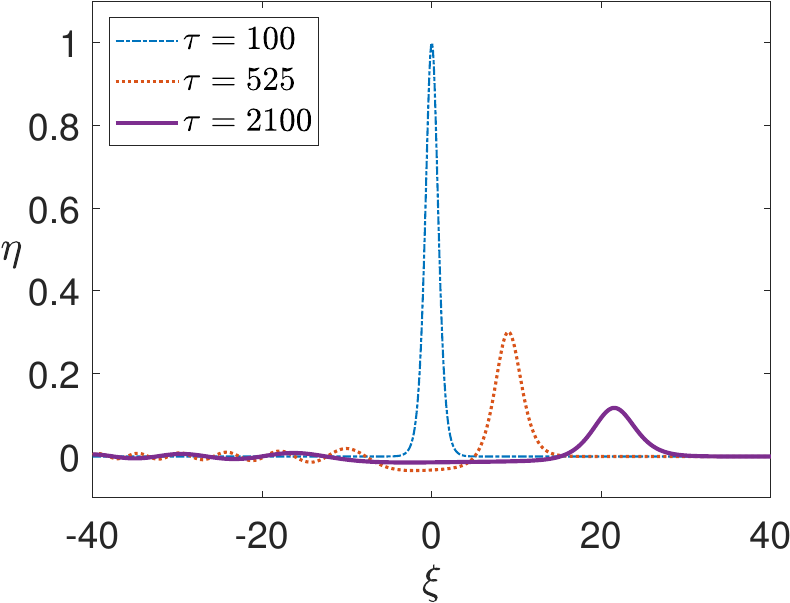}
\includegraphics[width = 8cm]{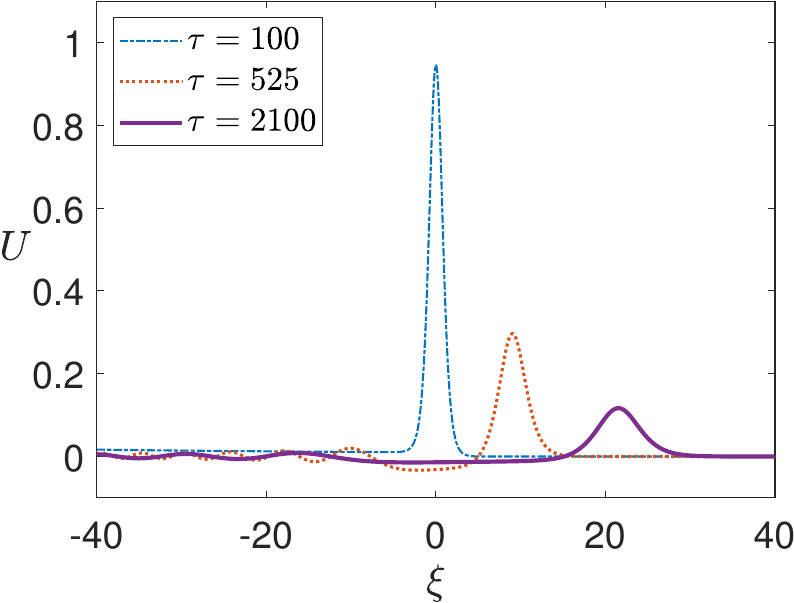}
\vspace{-0.3cm}
\caption{Boussinesq $\eta$ (left) and $U$ (right) solutions at varying $\tau$
for $\eps = 0.1$, $\tau \in [100,2100]$, $\xi \in [-40,40]$ (with the full computational domain $\xi \in [-90,60]$) and initial condition $\eta_0 = \sech^2(\xi)$. Physically, the initial condition a wave which is placed at the distance $r_0 = \taumin = 100$ from the origin and it then observed around distances $r\in \{ 525,2100 \}$. All visible waves are outward-propagating. }
\label{eps0_1_tauprofiles}
\end{figure}

\begin{figure}
\centering
\includegraphics[width = 8cm]{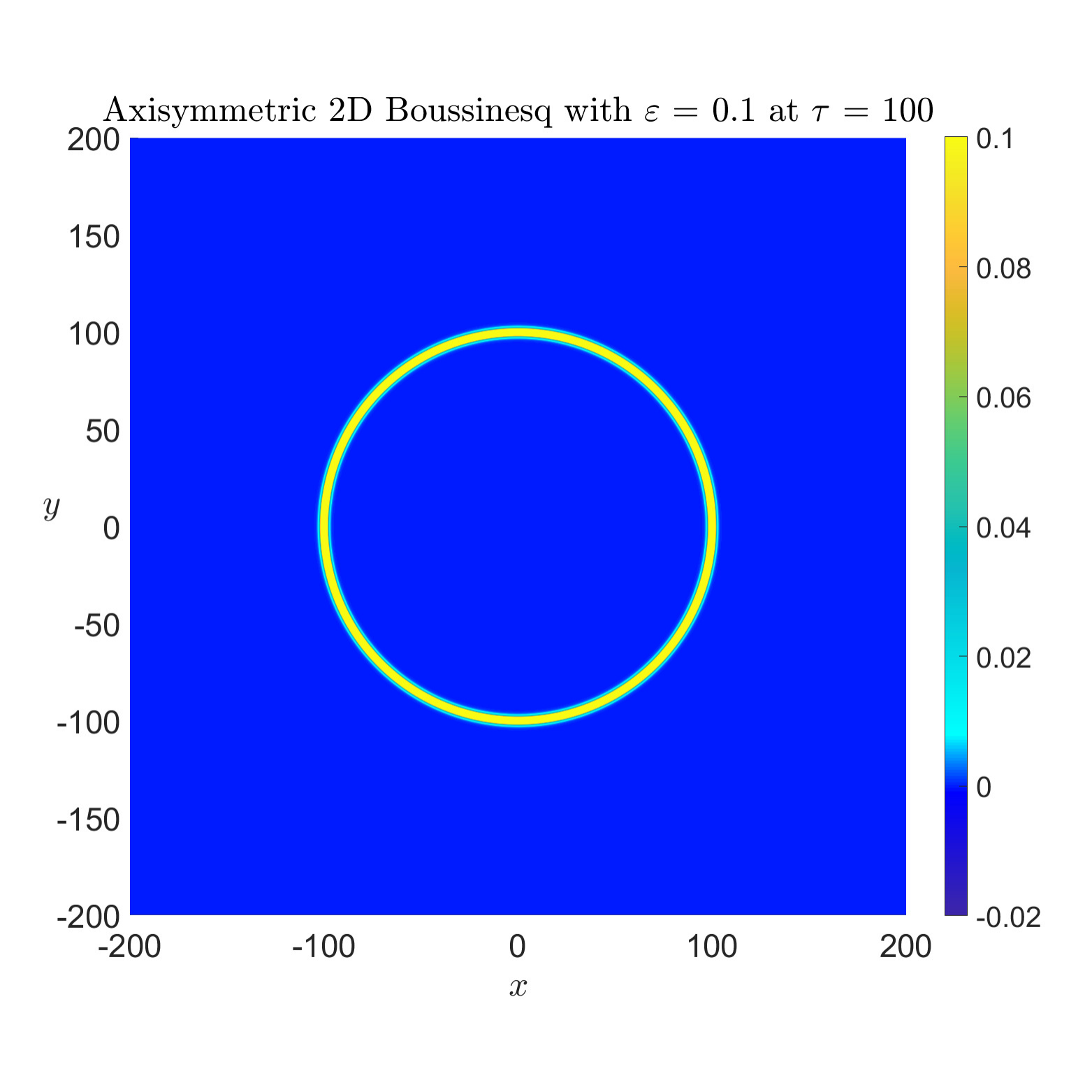}
\includegraphics[width = 8cm]{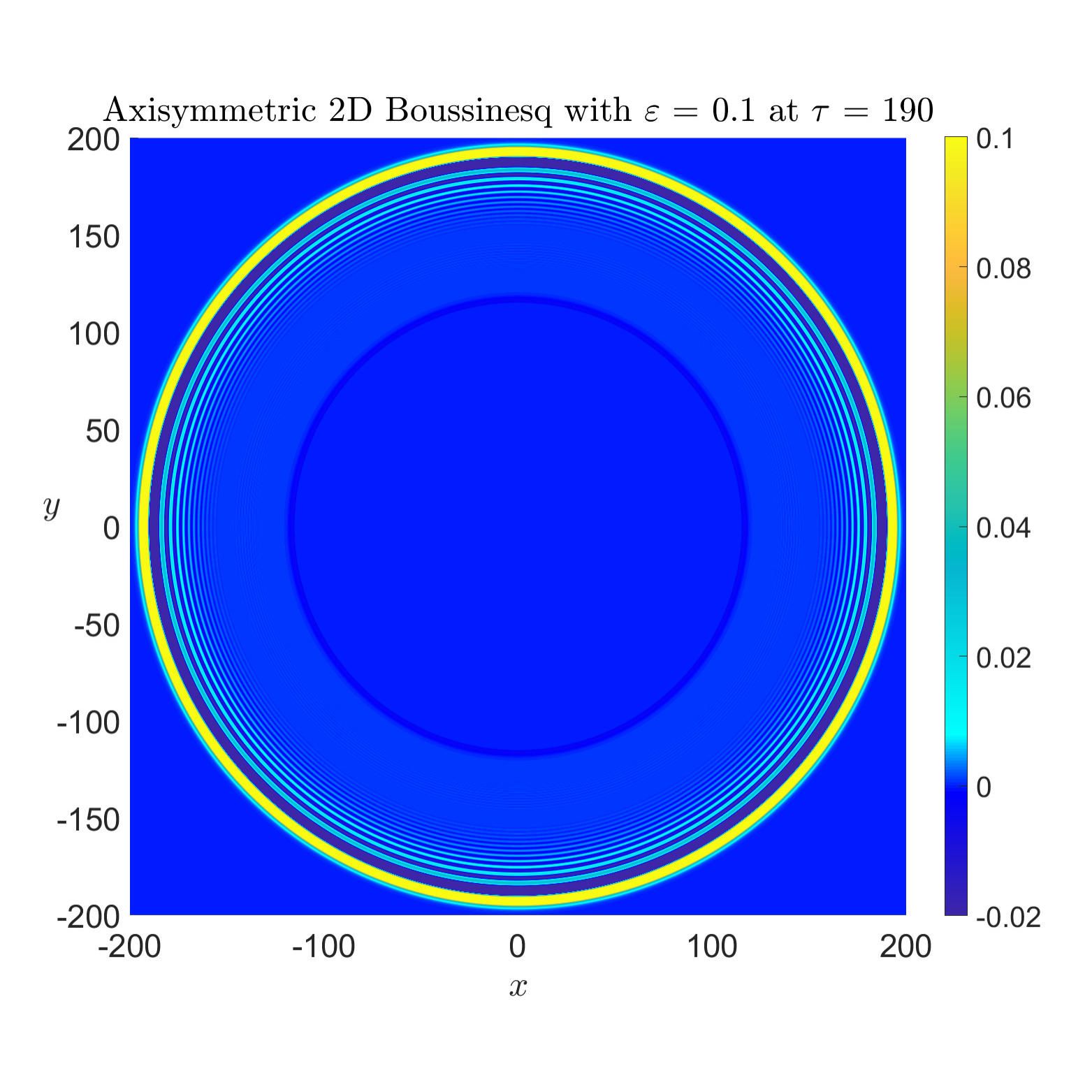}
\vspace{-0.8cm}
\caption{Top-down view of the $\eta$ solution to the axisymmetric 2D Boussinesq model for $\eps = 0.1$ at varying $\tau$. The initial condition is $\eta_0 = \sech^2\xi$ at $\tau = 100$ corresponding to a wave being placed at a distance $r_0=100$ from the origin. The shelf and oscillatory tail develop quite quickly. The suppression of waves near the boundaries of the computational domain is represented by the dark annulus behind the main ring. The modelled solution is contained between the dark annulus and the bright lead ring wave. }
\label{eps0_1_topdown}
\end{figure}

The initial condition has the form of a localised outward-propagating wave. The condition needs to be sufficiently accurate, since any waves that are inward-propagating in the fixed reference frame -- however small they are initially -- will grow due to cylindrical convergence, which may lead to numerical instability. We use the $O(\varepsilon)$ correction following from the uni-drectional model discussed in the next section in order to formulate a suitable initial condition.   As the wave propagates, a shelf and an oscillatory transition regions develop behind the lead pulse, both in the free-surface elevation and in the radial velocity, which can be seen in Figure \ref{eps0_1_tauprofiles}. All visible waves in this Figure \ref{eps0_1_tauprofiles} are outward-propagating, but the oscillatory wave train moves slower than the linear long waves and therefore it appears to be inward-propagating in this moving coordinate frame. The typical topdown view of the simulated concentric waves of the axisymmetric Boussinesq system is shown in Figure \ref{eps0_1_topdown}. The bright lead ring wave of elevation is followed by the dark ring wave of depression (the shelf), which is then connected to the undisturbed medium by the oscillatory transition region (a transient ring dispersive shock wave). The visible inner circle corresponds to the boundary of the effective computational domain, which is discussed in detail in Section \ref{sec:Num}.   \\

Instead of the weakly-nonlinear Boussinesq equations, one can adopt  long-wave models with higher-order nonlinear and dispersive terms, such as the higher-order 2D Boussinesq models:  Serre-Green-Naghdi (SGN) \cite{Se, GN, SG} or Matsuno (extended SGN) \cite{M1, M2} equations. However, it should be pointed out that the initial value problems for these higher-order models have been known to be ill-posed \cite{M1,C1} and, therefore, cannot be used for numerical simulations without a suitable regularisation.

\section{Extended cylindrical Korteweg-de Vries (ecKdV) equation}
\label{sec:cKdV}

We now look for solutions of the  Boussinesq system (\ref{Boussinesq2Dcylindrical}) in the form of  asymptotic multiple-scale expansions 
\begin{align}
\label{etatilde}
\Te & = \etazero + \eps \etaone + \eps^2 \etatwo + O(\eps^3), \\
\label{Utilde}
\TU & = \Uzero + \eps \Uone + \eps^2 \Utwo + O(\eps^3),
\end{align}
where all functions $\eta^{(i)},U^{(i)}$ depend on the characteristic fast variable $\xi = \Tr - \Tt$, and the slow radial variable $R = \eps \Tr$. 
At leading order, $O(1)$, we obtain
\begin{equation}
\label{BoussinesqO1}
\begin{cases}
-\etazero_\xi + \Uzero_\xi = 0 , \\[0.3cm]
-\Uzero_\xi + \etazero_\xi = 0,
\end{cases}
\end{equation}
which implies that
\begin{equation}
\label{Uone}
\Uzero = \etazero + f(R).
\end{equation}
We let $f(R) = 0$ under the assumption that all disturbances are produced only by the propagating waves. Physically this means we impose undisturbed conditions ahead of the outward-propagating ring waves. At a fixed $r$, before any waves reach this point (i.e. $t \to -\infty$ so that $\xi \to \infty$), both functions must satisfy $\Uzero,\etazero \to 0$.  \\

Next, at $O(\eps)$, we have
\begin{equation}
\label{BoussinesqO2}
\begin{cases}
-\etaone_\xi +\Uone_\xi + \Uzero_R + [\etazero \Uzero]_\xi + \dfrac{1}{R} \Uzero = 0, \\[0.3cm]
\etaone_\xi - \Uone_\xi + \etazero_R + \dfrac{1}{2}[\Uzero \Uzero]_\xi + \dfrac{1}{3}\Uzero_{\xi\xi\xi}= 0,
\end{cases}
\end{equation}
and here we add the two equations, and also substitute the result (\ref{Uone}) to obtain the cKdV equation:
\begin{equation}
\label{ckdv}
\etazero_R + \dfrac{3}{2}\etazero \etazero_\xi + \dfrac{1}{6}\etazero_{\xi\xi\xi} + \dfrac{1}{2R}\etazero = 0.
\end{equation}
From the first equation of (\ref{BoussinesqO2}), we find
\begin{equation}
\Uone_\xi = \etaone_\xi - \etazero_R - 2\etazero \etazero_\xi - \dfrac{1}{R}\etazero.
\end{equation}
Using the cKdV equation (\ref{ckdv}) and integrating over $\xi' \in (\xi, \infty)$, we obtain
\begin{equation}
\label{Utwo}
\Uone = \etaone - \dfrac{1}{4}\big[\etazero\big]^2 + \dfrac{1}{6}\etazero_{\xi\xi} - \dfrac{1}{2R}\phizero, \qquad \phizero = -\int_{\xi}^{\infty} \etazero(\xi',R) \ d\xi'. 
\end{equation}
Here, we implicitly assumed that the waves propagate into the unperturbed medium, which is a natural condition for the problems with localised initial data.\\

Finally, at $O(\eps^2)$, we have
\begin{equation}
\label{BoussinesqO3}
\begin{cases}
-\etatwo_\xi + \Utwo_\xi + \Uone_R + [\etazero\Uzero]_R + [\etazero\Uone]_\xi + [\etaone\Uzero]_\xi + \dfrac{1}{R}[\etazero\Uzero + \Uone]= 0 , \\[0.3cm]
\etatwo_\xi - \Utwo_\xi + \etaone_R + \dfrac{2}{3} \Uzero_{R\xi\xi} + \Uzero\Uzero_R + \dfrac{1}{3}\Uone_{\xi\xi\xi} + [\Uzero \Uone]_{\xi} + \dfrac{1}{3R}\Uone_{\xi\xi} = 0.
\end{cases}
\end{equation}

We add these two equations to eliminate $\etatwo,\Utwo$ terms, and use the following relations:
\begin{subequations}
\begin{align}
\Uzero & = \etazero, \\[0.3cm]
\Uone & = \etaone - \dfrac{1}{4}[\etazero]^2 + \dfrac{1}{6}\etazero_{\xi\xi} - \dfrac{1}{2R}\phizero, \\[0.3cm]
\etazero_R & = -\dfrac{3}{2}\etazero\etazero_{\xi} - \dfrac{1}{6}\etazero_{\xi\xi\xi} - \dfrac{1}{2R}\etazero, \\[0.3cm]
\etazero_{R\xi\xi} & = -\dfrac{9}{2} \etazero_{\xi} \etazero_{\xi\xi} -  \dfrac{3}{2} \etazero\etazero_{\xi\xi\xi} - \dfrac{1}{6} \etazero_{\xi\xi\xi\xi\xi} - \dfrac{1}{2R} \etazero_{\xi\xi}, \\[0.3cm]
\phizero_\xi & = \etazero, \\[0.3cm]
\phizero_R & = -\int_{\xi}^{\infty} \etazero_R (\xi',R) \ d\xi' = -\dfrac{3}{4}\big[ \etazero \big]^2 - \dfrac{1}{6}\etazero_{\xi\xi} - \dfrac{1}{2R}\phizero.
\end{align}
\end{subequations}
The resulting equation takes the form
\begin{align}
\label{eckdvO2}
\etaone_R + \dfrac{3}{2} \big[\etazero\etaone\big]_\xi + \dfrac{1}{6}\etaone_{\xi\xi\xi} + \dfrac{1}{2R}\etaone &- \dfrac{21}{8}\big[\etazero\big]^2\etazero_\xi - \dfrac{47}{24}\etazero_\xi\etazero_{\xi\xi} - \dfrac{3}{4}\etazero\etazero_{\xi\xi\xi} - \dfrac{1}{24}\etazero_{\xi\xi\xi\xi\xi} \nonumber \\[0.2cm]
& -\dfrac{1}{16R}\bigg[ 9\big[\etazero\big]^2 + 8\etazero_\xi\phizero \bigg] + \dfrac{1}{8R^2}\phizero = 0.
\end{align}
We now consider the truncated expansion (\ref{etatilde}) and let $\He = \etazero + \eps \etaone$. Multiplying equation (\ref{eckdvO2}) through by $\eps$, adding this to equation (\ref{ckdv}), and noting that
\begin{equation}
\He \He_\xi = \etazero\etazero_\xi + \eps\big[\etazero\etaone\big]_\xi + O(\eps^2),
\end{equation}
we obtain the extended cKdV (ecKdV) equation
\begin{align}
\label{eckdv}
\He_R + \dfrac{3}{2} \He\He_\xi + \dfrac{1}{6}\He_{\xi\xi\xi} + \dfrac{1}{2R}\He &-\eps \bigg( \dfrac{21}{8}\He^2 \He_\xi + \dfrac{47}{24}\He_\xi \He_{\xi\xi} + \dfrac{3}{4}\He\He_{\xi\xi\xi} + \dfrac{1}{24}\He_{\xi\xi\xi\xi\xi} +\dfrac{1}{16R}\bigg[ 9\He^2 + 8\He_\xi \Hphi \bigg] - \dfrac{1}{8R^2}\Hphi \bigg) = 0,
\end{align}
where we have omitted any $O(\eps^2)$ terms and defined
\begin{equation}
\Hphi = -\int_{\xi}^{\infty} \He(\xi',R) \ d\xi'.
\end{equation}
\\
It should be noted that  the coefficients of the ecKdV equation depend on its parent long wave model. 
In the Appendix,  
 a slow radius version of the ecKdV equation is derived from Matsuno's 2D extended SGN model \cite{M1}
 in the form
\begin{align}
\label{extendedCKDVrelationMATSUNO1}
\He_R + \dfrac{3}{2} \He\He_\xi + \dfrac{1}{6}\He_{\xi\xi\xi} + \dfrac{1}{2R}\He &-\eps \bigg( \dfrac{21}{8}\He^2 \He_\xi + \dfrac{7}{12}\He\He_{\xi\xi\xi} + \dfrac{31}{24}\He_\xi \He_{\xi\xi} + \dfrac{11}{360} \He_{\xi\xi\xi\xi\xi} +\dfrac{1}{16R}\bigg[ 9\He^2 + 8\He_\xi \Hphi \bigg] - \dfrac{1}{8R^2}\Hphi \bigg) = 0.
\end{align}
The difference in nonlinearity and dispersive coefficients is clear when comparing this to (\ref{eckdv}). 
Also, in the Appendix we show that the 2D SGN equations \cite{Se,GN,SG} yield the extended cKdV equation in the form
\begin{equation}
\label{GNeckdv}
\He_R + \dfrac{3}{2} \He\He_\xi + \dfrac{1}{6}\He_{\xi\xi\xi} + \dfrac{1}{2R}\He -\eps \bigg( \dfrac{21}{8}\He^2 \He_\xi + \dfrac{7}{12}\He\He_{\xi\xi\xi} + \dfrac{31}{24}\He_\xi \He_{\xi\xi} + \dfrac{1}{24}\He_{\xi\xi\xi\xi\xi} +\dfrac{1}{16R}\bigg[ 9\He^2 + 8\He_\xi \Hphi \bigg] - \dfrac{1}{8R^2}\Hphi \bigg) = 0,
\end{equation} 
 which differs from \eqref{extendedCKDVrelationMATSUNO1} only in the coefficient of the higher-order dispersive term $\He_{\xi\xi\xi\xi\xi}$, while the difference with the ecKdV equation \eqref{eckdv} derived from the 2D Boussinesq system is much more significant.
Thus, all three of these models have their own ecKdV equations which describe outward-propagating surface ring waves. We also derived similar versions for the inward-propagating waves (see the Appendix). \\

It is also instructive to compare the slow radius $R = \eps \tilde{r}$ version of the extended cKdV equation \eqref{extendedCKDVrelationMATSUNO1} derived from Matsuno's model \cite{M1} to a slow time  $T = \eps \tilde{t}$ version derived by Horikis et al. \cite{HFMS1, HFMS2} from the potential formulation of the Euler equations.  Using the change of variable $R = T + \eps \xi$, followed by asymptotically equivalent transformations by virtue of the derived equation (\ref{extendedCKDVrelationMATSUNO1}), we obtain

\begin{eqnarray}
&&\He_T +  \dfrac{3}{2} \He \He_{\xi} + \dfrac{1}{6} \He_{\xi\xi\xi} + \dfrac{1}{2T} \He +
  \eps \left [-\dfrac{3}{8} \He^2 \He_{\xi} + \dfrac{5}{12}  \He \He_{\xi\xi\xi}+ \dfrac{23}{24} \He_{\xi} \He_{\xi\xi}  + \dfrac{19}{360} \He_{\xi\xi\xi\xi\xi} \right . \nonumber \\  
&&\left . + \dfrac{1}{T} \left( \dfrac{3}{16} \He^2 + \dfrac{1}{4} \He_{\xi\xi} - \dfrac{1}{2} \Hphi \He_{\xi}\right ) + \dfrac{1}{T^2} \left( \dfrac{1}{8}\Hphi - \dfrac{1}{2} \xi \He \right) \right] = 0,\ \mbox{where}\
 \Hphi = -\int_{\xi}^{\infty} \He(\xi',T) \ d\xi'.
\end{eqnarray}
The model is asymptotically equivalent to that in \cite{HFMS1, HFMS2} provided that $\int_{\xi}^\infty \hat \phi_T d \xi$ is bounded,
up to the definition of $\Hphi$ which in their version is defined to be
\begin{equation*}
\Hphi = \int_{0}^{\xi} \He(\xi',T) \ d\xi',
\end{equation*}
corresponding to the absence of waves at the origin. We believe that it is more natural to assume that the waves propagate into an unperturbed medium, and this is what has led to the form of the non-local term in our version of the model. Also, a slow radius form of the equation is a more natural choice from the viewpoint of experiments and observations, since it is unlikely that one will know the state of the entire fluid at some moment of time, while the state at some distance around the origin can be registered by buoys (see \cite{RS, G}). Note that unlike the leading order cKdV equation, the slow-time and the slow-radius versions of the extended cKdV equation are significantly different.

\section{Numerical modelling of outward-propagating concentric waves}
\label{sec:Num}

\begin{figure}
\centering
\includegraphics[width = 11cm]{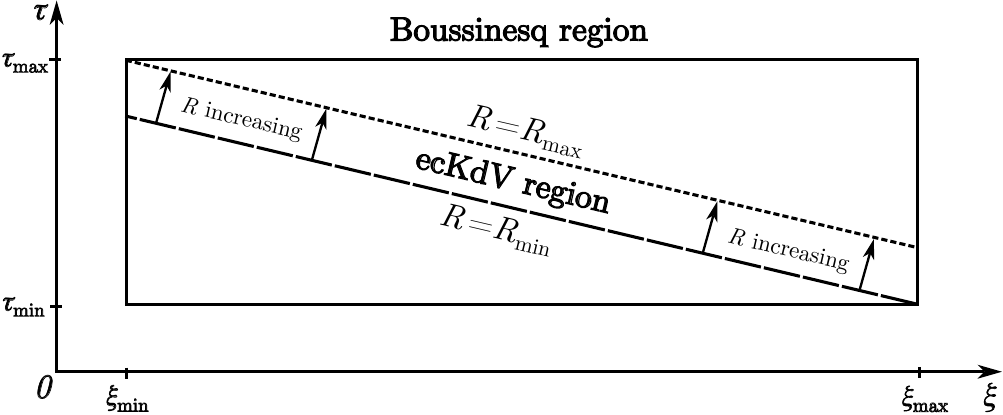}

\caption{Integration domains for the axisymmetric Boussinesq system in the moving reference frame  \eqref{Boussinesq2DcylindricalMoving} and the ecKdV model \eqref{eckdv}.}
\label{bouss_to_ckdv_domain}
\end{figure}

In this section, we discuss the practical aspects of using the reduced cKdV-type models to accelerate numerical modelling of outward-propagating concentric waves for a given initial condition. We experiment with localised initial conditions placed away from the origin and generating an outward-propagating ring wave, but the approach can be used with other initial data and other parent systems. \\

Numerical runs for the axisymmetric Boussinesq system have been initialised at $\tau = \taumin$ with initial profiles $\He_0, \HU_0$ over the domain 
$$\xi \in [\ximin,\ximax ].$$
 After integrating this model up to $\tau = \taumax$, we extract initial data for cKdV and ecKdV models at fixed  
 $$R = \Rmin = \eps (\ximax + \taumin)$$
  and simulate these up to 
  $$R = \Rmax = \eps (\ximin + \taumax).$$
\\
The cKdV equation is a far-field model and,  therefore, $\Rmin$ and $\Rmax$ cannot be too small. In our current modelling, we do not have a fully defined solution across our entire $\xi$-domain if we impose the initial condition to be at an $R$ value smaller than $\Rmin$, which is clarified in Figure \ref{bouss_to_ckdv_domain}. Indeed, the axisymmetric Boussinesq system \eqref{Boussinesq2Dcylindrical} is integrated over the region $\Omega = [\ximin,\ximax]\times[\taumin,\taumax]$. The initial data for cKdV-type models must be extracted at a particular $R$ where $R = \eps \tilde{r} = \eps(\xi + \tau)$ (or $\tau = R/\eps - \xi$) since $\xi = \tilde{r}-\tau$. In $(\xi,\tau)$-space the curves corresponding to constant $R$ are straight lines which must fit entirely on $\Omega$ for the $\eta$ profiles to be extracted and compared over the entire $\xi$-region. The minimum such $R=\Rmin$ corresponds to a straight line passing through $(\ximax,\taumin)$ (represented by dashed line) hence giving $\Rmin = \eps(\ximax + \taumin)$, and likewise $\Rmax = \eps(\ximin + \taumax)$ is a relation satisfied when passing through $(\ximin,\taumax)$ instead (represented by dotted line).
This explains the choice of the $\Rmin$ and $\Rmax$ values used in the majority of our numerical runs. However, we will also consider simulations over the region much closer to the origin, at the expense of describing the solution in a smaller $\xi$ domain.
\\

We then compare the cKdV and ecKdV solutions at $R = \Rhat$, where 
  $$\Rhat \in [\Rmin,\Rmax ],$$  
    and also compare with the full Boussinesq solution at this value. We shall denote the approximate solutions to $\eta$ given by cKdV and ecKdV equations using $\He_{\text{ckdv}}$ and $\He_{\text{eckdv}}$, respectively. Exact Boussinesq solutions $\eta$ will be denoted by $\He_{\text{bouss}}$. \\
    
All three models are solved using pseudospectral schemes with $N$ nodes for $\xi$ derivatives and the 4th-order Runge-Kutta scheme for $\tau$ and $R$ derivatives with step-sizes $\Delta \tau$ and $\Delta R$, respectively. The parameters are fixed as follows:
\begin{equation*}
(\ximin,\ximax) = (-90,60), \quad (\taumin, \taumax) = (100,2100), \quad N=3 \times 2^9,\quad \Delta \tau = 5 \times 10^{-2},\quad  \Delta R = 10^{-4}.
\end{equation*} \\

The initial data for the Boussinesq system is given by
\begin{equation}
\label{eta0U0}
\eta_0 = A\sech^2 \lambda \xi, \quad U_0 = \Uzero + \eps \Uone = \eta_0 + \dfrac{1}{2(\xi + \taumin)} \int_{\xi}^{\ximax} \eta_0 \ d\xi - \eps \left( \dfrac{1}{4} \eta_0^2 - \dfrac{1}{6} \eta_{0\xi\xi} \right),
\end{equation}
where $A,\lambda>0$ define the amplitude and width of our initial pulse, respectively. The weakly-nonlinear approximation for $U_0$ is based on the relations for $\Uzero,\Uone$ as found in (\ref{Uone}), (\ref{Utwo}). We note that it was important to use this improved initial condition to simulate the outward-propagating waves. Indeed, it reduces the amplitude of the inward-propagating wave, which inevitably starts growing due to cylindrical convergence and may lead to instability. \\


 \begin{figure}
\centering
\includegraphics[width = 8cm]{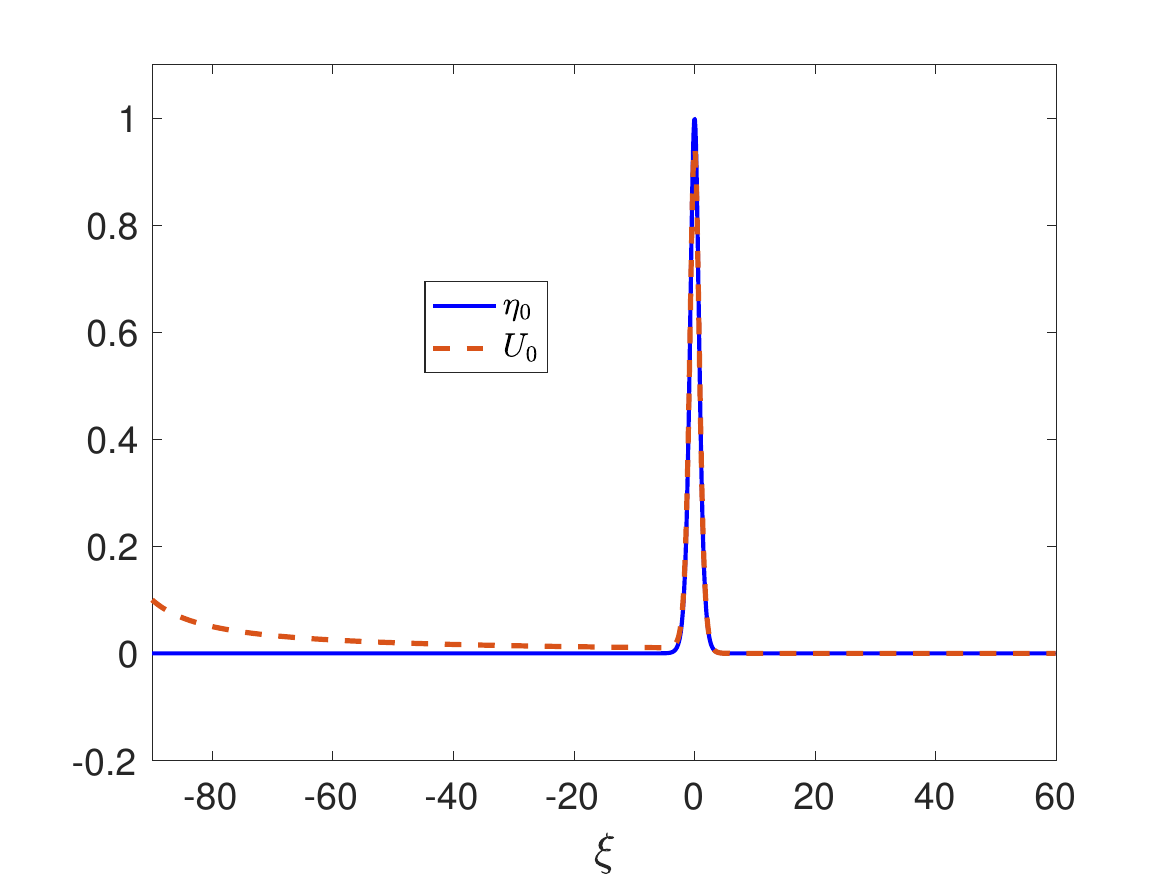}
\includegraphics[width = 8cm]{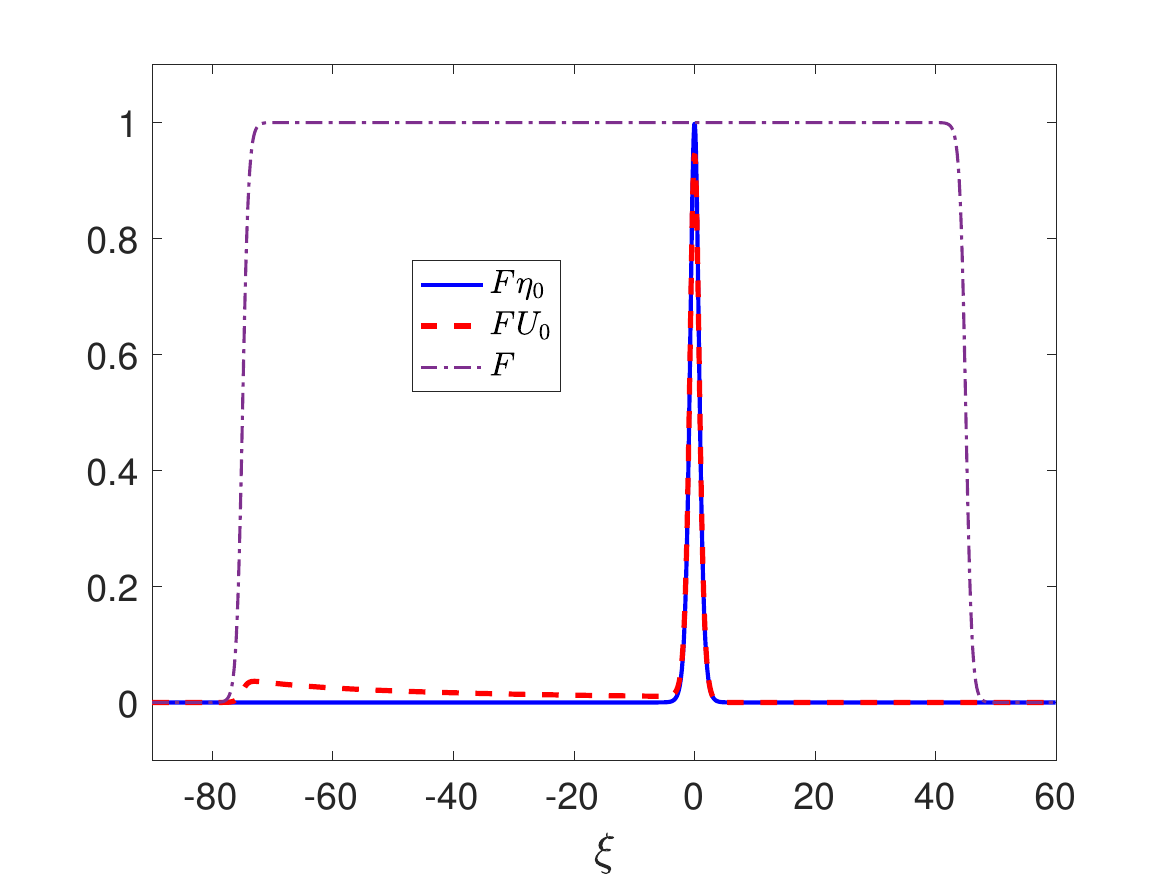}
\caption{Initial conditions $\eta_0,U_0$ are modified using multiplication by the function $F$ to bring them to zero at the ends of the interval. The modified functions $F\eta_0,F U_0$ can be treated as periodic over the computational domain. 
The modification of $\eta_0,U_0$ are shown for $A = \lambda = 1$ and $\eps = 0.1$.}
\label{BoussIC_Filtering}
\end{figure}

In the numerical work, we employ MATLAB \cite{Matlab} to solve the moving-frame, axisymmetric $(\xi,\tau)$-Boussinesq system \eqref{Boussinesq2DcylindricalMoving} as well as the related cKdV and ecKdV models in the $(\xi,R)$-space \eqref{ckdv} and \eqref{eckdv} for the variable $\He$ (equal to $\etazero + \eps \etaone$). We solve these models over spatial domain $\xi \in [ \ximin, \ximax]$ and ``temporal" domains $\tau \in [\taumin, \taumax]$ and $R \in [\Rmin,\Rmax]$. The spatial $\xi$-derivatives in all cases are computed using the fast Fourier transform algorithm provided by MATLAB. 
The Boussinesq system has two initial conditions $\hat{\eta}_0,\hat{U}_0$ producing the outputs $\He,\HU$ at a fixed time $\tau$. The initial data given by \eqref{eta0U0} was modified near the ends of the $\xi$ domain in order to maintain periodicity. Indeed, using Fourier transforms requires the dependent variables $\He,\HU$ to be periodic on the specified $\xi$ domain. However, this condition is violated by the way $U_0$ depends on $\eta_0$. To address this issue, we introduce a function $F$ to modify this initial data $\eta_0,U_0$ into periodic data $\hat{\eta}_0,\hat{U}_0$ across our $\xi$ domain. Here $F$ is defined as
\begin{equation}
\label{F_suppression}
F = F(\xi) = \dfrac{1}{2} \left( \tanh \bigg[ \kappa_F \left( \xi - \ximin - \dfrac{ \xi_{\text{span}} }{10} \right) \bigg] + \tanh \bigg[ \kappa_F \left( \xi - \ximax + \dfrac{ \xi_{\text{span}} }{10} \right) \bigg] \right),
\end{equation} 
where $\xi_{\text{span}} = \ximax - \ximin$ and $\kappa_F > 0$ is a constant. Then we can consider the periodic data 
$$(\He_0,\HU_0) = (F \eta_0, FU_0)$$
 which coincides with $( \eta_0, U_0)$ over most of the domain except near the boundaries $\xi \in \{ \ximin,\ximax \}$ where the function $F$ (filter)  brings the solutions to $0$. The parameter $\kappa_F$ denotes the rate of the suppression which we set as $\kappa_F = 1.5$ in the majority of our experiments. Figure \ref{BoussIC_Filtering} shows this modification. \\
 
To numerically integrate in time, we apply the 4th-order Runge-Kutta (RK4) scheme with timestep $h \in \{ \Delta \tau, \Delta R \}$. The RK4 scheme has been adjusted to include multiplication by $F$ at every timestep. Hence, if the Cauchy problem is to solve $\eta_{\Tt} = \psi(\Tt,\eta)$ then the RK4 scheme for integrating in $\Tt$ is defined as:
\begin{align}
\eta(\Tt_{n+1},\xi) = F \cdot \bigg[ \eta(\Tt_n,\xi) + \dfrac{h}{6}(k_1 + k_2 + k_3 + k_4) \bigg] , \qquad \Tt_{n+1} = \Tt_n + h, \qquad \Tt \in \{ \tau,R \},
\end{align}
where
\begin{subequations}
\begin{align}
k_1 & = \psi (\Tt_n,\eta_n ), \\
k_2 & = \psi \bigg(\Tt_n + \frac{h}{2},\eta_n + h\frac{k_1}{2} \bigg), \\
k_3 & = \psi \bigg(\Tt_n + \frac{h}{2},\eta_n + h\frac{k_2}{2}\bigg), \\
k_4 & = \psi (\Tt_n + h,\eta_n + hk_3).
\end{align}
\end{subequations}

Hence, any waves which get too close to the boundaries of the $\xi$ domain are suppressed. \\

For the cKdV-type models, the numerical approach is somewhat different. Firstly, it is important to note that these models are stiff requiring, with the direct head-on approach, $\Delta R \sim O(10^{-5}) - O(10^{-6})$ which results in simulations almost as expensive as the parent axisymmetric Boussinesq system in our present runs. Stiffness is a characteristic property of some PDEs requiring the use of additional tools. There are several methods one can apply for stiff PDEs, including the integrating factor method (e.g. \cite{KT}). Here, we adapt the  integrating factor method for the KdV equation to our cKdV-type models. To do so, we write the ecKdV equation in the form
\begin{align}
\label{ckdv_linear_nonlinear}
\He_R + \gamma_2 \He_{\xi\xi\xi} - \eps \gamma_7 \He_{\xi\xi\xi\xi\xi} = \underbrace{- \gamma_1 \He\He_\xi - \dfrac{\gamma_3}{2R}\He + \eps \bigg( \gamma_4 \He^2 \He_\xi + \gamma_5 \He_\xi \He_{\xi\xi} + \gamma_6 \He\He_{\xi\xi\xi} +\dfrac{1}{R}\bigg[ \gamma_8 \He^2 + \gamma_9 \He_\xi \Hphi \bigg] + \dfrac{\gamma_{10}}{R^2}\Hphi \bigg)}_{= \mathcal{N}(R,\He)},
\end{align}
where we have separated linear terms from nonlinear and non-local terms (denoted by $\mathcal{N}$). We now apply the Fourier transform $\F$ to both sides using the notation $\He^* = \F[\He]$ and the property $\F[\He_{n\xi}] = (\im k)^n \He^*$ (where the subscript $n\xi$ denotes the order of the $\xi$ derivative) giving us
\begin{align}
\He^*_R + \Lambda \He^* = \F[\mathcal{N}(R,\F^{-1}[\He^*])], \quad \text{where} \quad \Lambda = - \im k^3 (\gamma_2 + \eps k^2 \gamma_7),
\end{align}
where $\F^{-1}$ denotes the inverse Fourier transform. Next, we define $e^{\Lambda R}$ as our integrating factor allowing us to introduce the variable $\nu = e^{\Lambda R} \He^*$ and rewrite the previous equation in the following form
\begin{align}
\label{ckdv_IF_nosponge}
\nu_R = e^{\Lambda R} \F[\mathcal{N}(R,\F^{-1}[e^{-\Lambda R} \nu])],
\end{align}
with the initial data
\begin{align}
\label{ckdv_IF_nosponge}
\nu(\Rmin,\xi) = e^{\Lambda \Rmin} \F[\He(\Rmin,\xi)].
\end{align}
The term $\He(\Rmin,\xi)$ denotes the initial data which we pass to the cKdV and ecKdV models from our axisymmetric Boussinesq runs at $R = \Rmin$. \\

We want to integrate \eqref{ckdv_IF_nosponge} however, as $\nu$ is a Fourier space variable, we are unable to simply use $F$ as defined by \eqref{F_suppression} at each time-step of the RK4 algorithm. To get around this, we turn off spatial suppression in RK4 by setting $F=1$ and instead introduce the sponge layer $s$ defined as
\begin{equation}
s = s(\xi) = \sigma \left[ 1 - \dfrac{1}{2} \left( \tanh \bigg[ \kappa_s \left( \xi - \ximin - \dfrac{ \xi_{\text{span}} }{10} \right) \bigg] + \tanh \bigg[ \kappa_s \left( \xi - \ximax + \dfrac{ \xi_{\text{span}} }{10} \right) \bigg] \right) \right],
\end{equation}
which behaves like $0$ in the middle of the domain and activates to strength $\sigma$ and rate $\kappa_s$ near the domain boundaries. The purpose of this sponge layer is to absorb any waves which get too close to the boundary. In our simulations we set $\kappa_s = \kappa_F = 1.5$ and $\sigma = 750$ which ensures that no waves cross the boundary. The choice for $\sigma$ ensures that waves corresponding to large $\eps = 0.5$ do not pass the boundary. However, it can be reduced when considering smaller $\eps$ waves. The sponge layer is inserted into the cKdV-type models by making the change $\He_R \mapsto \He_R + s\He$ which means an additional term to occur alongside $\mathcal{N}$. Thus, \eqref{ckdv_IF_nosponge} is adjusted to
\begin{align}
\label{ckdv_IF_sponge}
\nu_R = e^{\Lambda R} \F[\mathcal{N}(R,\F^{-1}[e^{-\Lambda R} \nu]) - s \F^{-1}[e^{-\Lambda R} \nu] ]. 
\end{align}
The solutions can then be easily inverted back to the real space by virtue of the transformation $\He = \F^{-1}[e^{-\Lambda R}\nu]$. \\

Lastly, the non-local term $\Hphi$ is computed numerically via composite Boole's rule \cite{Boole} which is a high-order, finite-difference numerical integration scheme. The aim is to use Boole's rule on the majority of the $\xi$ domain followed by lower-order methods at the right boundary as our quantity of nodes decreases. We define the mesh points
$$\xi_n := \ximin + (n-1) \Delta \xi, \qquad \He_n:=\He(R,\xi_n),$$
where $n=1,2,\ldots,N$. (Note that due to periodicity in our numerical implementation we exclude $\xi_{N+1} = \ximax$ whereby $\He_{N+1} = \He_1$.) Since Boole's rule requires the number of strips to be a multiple of $4$, we apply it to the first $4K$ strips, where $K = \lfloor (N+1-n)/4 \rfloor$, and use a lower-order method for the remaining strips. Then the non-local term at the mesh points is expressed as follows:
\begin{align}
\label{booles_rule}
\Hphi_n & := \Hphi(R,\xi_n) = \int_{\xi_n}^{\xi_{N+1}} \He \ d\xi = \int_{\xi_n}^{\xi_{n+4K+1}} \He \ d\xi + \int_{\xi_{n+4K+1}}^{\xi_{N+1}} \He \ d\xi \nonumber \\
& = \dfrac{2}{45} \Delta \xi \bigg[ 7(\He_n + \eta_{n+4K+1}) + 32 \sum_{i \in I_1} \He_i + 12 \sum_{i \in I_2} \He_i + 14 \sum_{i \in I_3} \He_i \bigg] + \int_{\xi_{n+4K+1}}^{\xi_{N+1}} \He \ d\xi + (\text{error}),
\end{align}
where
\begin{subequations}
\begin{align}
I_1 & = \{ n+1 , n+3 , n+5 , \ldots , n + 4K + 1 \},\\
I_2 & = \{ n+2 , n+6 , n+10 , \ldots , n + 4K \}, \\
I_3 & = \{ n+4 , n+8 , n+12 , \ldots , n + 4K - 2 \}.
\end{align}
\end{subequations}

The error term in this approximation is at least $O(\Delta\xi^7)$ (which is the error for Boole's rule). However, if $N+1-n \not\equiv 0 \pmod{4}$ then the error is higher as we need to use a lower-order method for the remaining strips. With three strips leftover we use Simpson's 3/8 rule with error $O(\Delta\xi^5)$. With two strips leftover we use Simpson's 1/3 rule having the same error order $O(\Delta\xi^5)$, and when we have only one strip leftover we approximate the area with a trapezium giving error $O(\Delta\xi^3)$. This way of calculating our integrals based on number of available strips (as we take larger $\xi_n$ to be the lower bound) allows us to calculate the non-local term $\Hphi$ to a high precision in the region away from the boundaries where waves are not being suppressed, with the lower-order methods being employed only at the right-most boundary of the domain, where the solution is exponentially small in size and variation due to wave suppression.
 \\

This choice of numerical schemes allow us to solve a typically very stiff PDE (the ecKdV) significantly more quickly by permitting larger time-steps $\Delta R$. In the numerical simulation for waves under the parameters 
$$\eps = 0.1, \quad [\ximin,\ximax] = [-90,60], \quad [\taumin,\taumax] = [100,2100], \quad N = 3 \times 2^9,$$
the time-step $\Delta \tau$ for the axisymmetric 2D Boussinesq system could not be taken larger than $5 \times 10^{-2}$ which can be validated via the linear stability analysis. This resulted in Boussinesq simulation taking $11.6$ hours to fully complete. These Boussinesq computations (along with the cKdV-type model computations as discussed later) were performed on a single machine with the 16-core AMD Ryzen™ 9 5950X processing unit. The 3D surface of $\He$ solutions for this Boussinesq simulation can be approximated via the cKdV-type models by taking the following steps:
\begin{enumerate}
\item Simulate the Boussinesq model from $\tau = \taumin$ to $\tau = \Rmin/\eps - \ximin$ where $\Rmin = \eps(\ximax + \taumin)$.
\item Extract the $\He$ profile from Boussinesq solution at $R=\Rmin$, and omit Boussinesq data in the region $R > \Rmin$.
\item Simulate cKdV-type models from $R = \Rmin$ up to $R = \eps(\ximax + \taumax)$ using the extracted profile as initial data.
\item Transform the cKdV surface of $\He$ solutions from $(R,\xi)$-space to $(\tau,\xi)$-space and append this surface to the adjusted Boussinesq surface resulting from step 2. Omit any data in the region $\tau > \taumax$.
\end{enumerate}
This approach yields exact Boussinesq solutions in the $(\tau,\xi)$-region where $R<\Rmin$ and cKdV-approximated solutions where $R>\Rmin$. Step 1 takes 50 minutes to complete. Step 3 can be done with, or without, the integrating factor. Without it the ecKdV is constrained to have small time-step $\Delta R = 10^{-5}$ resulting in step 3 taking 16.6 minutes via the cKdV, or 6.6 hours via the ecKdV. When the integrating factor is involved, we can take $\Delta R = 10^{-3}$ which means step 3 is completed in 3 minutes via the cKdV, or 40 minutes via the ecKdV. This indicates a substantial decrease in computation time when using the integrating factor, overall reducing the time taken to approximate Boussinesq $\He$ surface from 11.6 hours to just 1.5 hours using the ecKdV model. 
In the present runs, aliasing errors were negligible due to the use of a sufficiently large number of harmonics. Computations with a smaller number of harmonics with de-aliasing \cite{O, D} could be tried to reduce the computational time for all models.
Finally, we note that higher-order systems such as SGN or Matsuno's models are significantly more challenging computationally compared to the Boussinesq system whereas their asymptotic models at ecKdV level differ merely in some  coefficients which means that improvement in computation time and effort can be expected to be even greater.

\section{Asymptotic solution for cKdV equation vs numerical simulations}
\label{sec:Num1}

In this section, we first examine validity of the cKdV equation and its asymptotic solution obtained by  Johnson \cite{J2} 
in comparison with the solution of the Boussinesq system.  Johnson considers the following initial-value problem:
\begin{eqnarray}
&& \eta_R + \dfrac{3}{2}\eta \etazero_\xi + \dfrac{1}{6}\eta_{\xi\xi\xi} + \dfrac{1}{2R}\eta = 0, \\
&&\eta(\Rmin,\xi) = A \sech^2 \left( \dfrac{\sqrt{3A}}{2}\xi \right),
\end{eqnarray}
assuming that the initial condition is placed sufficiently far from the origin so that the cylindrical divergence term can be treated as a perturbation of the KdV solitary wave, defining the initial condition. \\

The approximate solution consists of three wave components: primary wave, shelf, and transition region back to the undisturbed state. To give the full description, we first introduce the following new variables
\begin{align}
\alpha = \Rmin^{-1}, \ \ X = \alpha R, \ \ \mathcal{T} = \dfrac{1}{2} \sqrt{3 A X^{-2/3}} , \ \ f(X) = \dfrac{3}{2} A ( X^{1/3} - 1 ), \\
\Theta = \xi - \alpha^{-1} f(X), \qquad \sbar = \sech(\mathcal{T} \Theta), \qquad \tbar = \tanh(\mathcal{T} \Theta).
\end{align}

The primary wave is then given by
\begin{equation}
\eta_{\text{primary}} = \eta_1 + \alpha \eta_2,
\end{equation}
 where
\begin{equation}
\eta_1  = A X^{-2/3} \sbar^2, \qquad \eta_2  = \frac{2}{3} \frac{X^{-2/3}}{\sqrt{3A}} \bigg[ -1 + \tbar + (3+2\mathcal{T} \Theta)\sbar^2  - \bigg( \frac{35}{12} + 3\mathcal{T} \Theta + \mathcal{T}^2 \Theta^2 \bigg) \tbar \sbar^2 \bigg].
\end{equation}

The shelf has the form
\begin{equation}
\eta_{\text{shelf}}  = \eta_1 + \alpha F_1  , \ \mbox{where} \ F_1 = -\dfrac{4}{3} (3A + 2 \alpha \xi)^{-1/2}.
\end{equation}

The oscillatory region is given by
\begin{equation}
\eta_{\text{oscill}}  = \eta_1 + \alpha F_2  , \  \mbox{where}  \  F_2  = -\dfrac{4}{3} (3A)^{-1/2} \bigg( 1 - \displaystyle \int_{\xi (2\alpha / X)^{1/3}}^{\infty} \mathrm{Ai}(\xi') \ d\xi' \bigg)
\end{equation}
 and $\mathrm{Ai}$ is the Airy function. Together these three components describe the cKdV evolution of a KdV soliton initial condition in $(R,\xi)$-space. The small parameter $\alpha$ depends on where we place the initial data, $\Rmin$, with the asymptotics being better with smaller $\alpha$. \\

To compare these asymptotics to the numerical solutions of the cKdV and Boussinesq models, we convert away from $(R,\xi)$-space to $(T,\xi)$-space in the same way as we did to compare against the work by Horikis et al. (2021) \cite{HFMS1}. At leading order, the cKdV equation retains its form with $T$ playing the role of $R$. This allows us to use the same initial data for Johnson's approximate solution,  the cKdV equation, and the Boussinesq system. In other words, we redefine Johnson's problem to be  the slow time problem,
and replace the $R$ with $T$ in Johnson's solution above. 
\\

\begin{figure}
\centering
\includegraphics[width = 8cm]{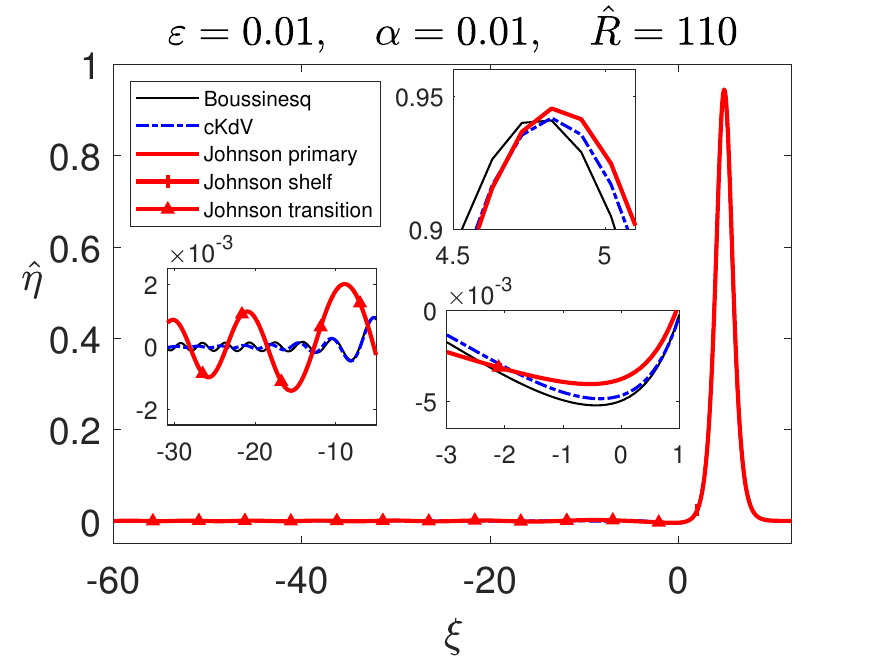}
\includegraphics[width = 8cm]{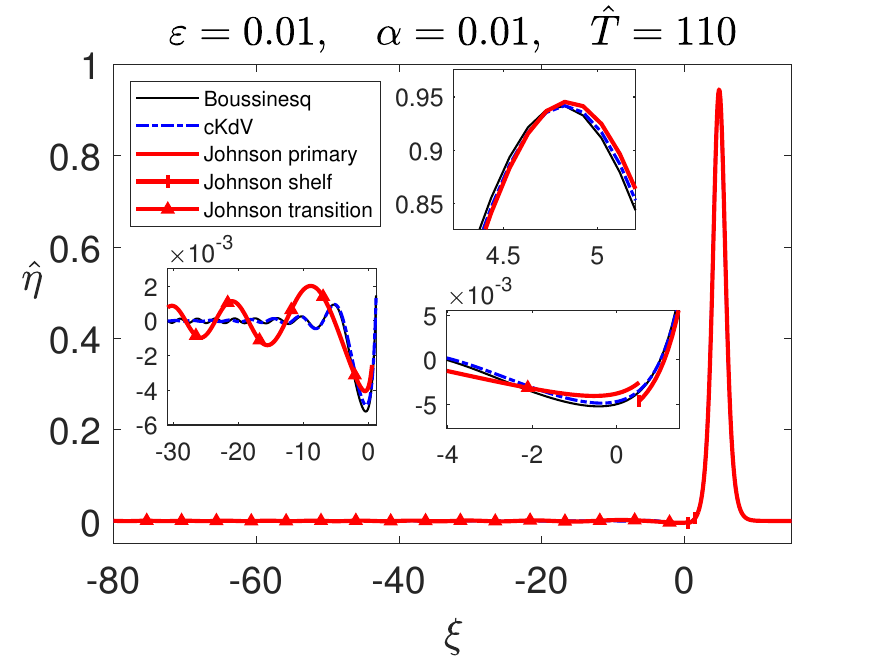}
\includegraphics[width = 8cm]{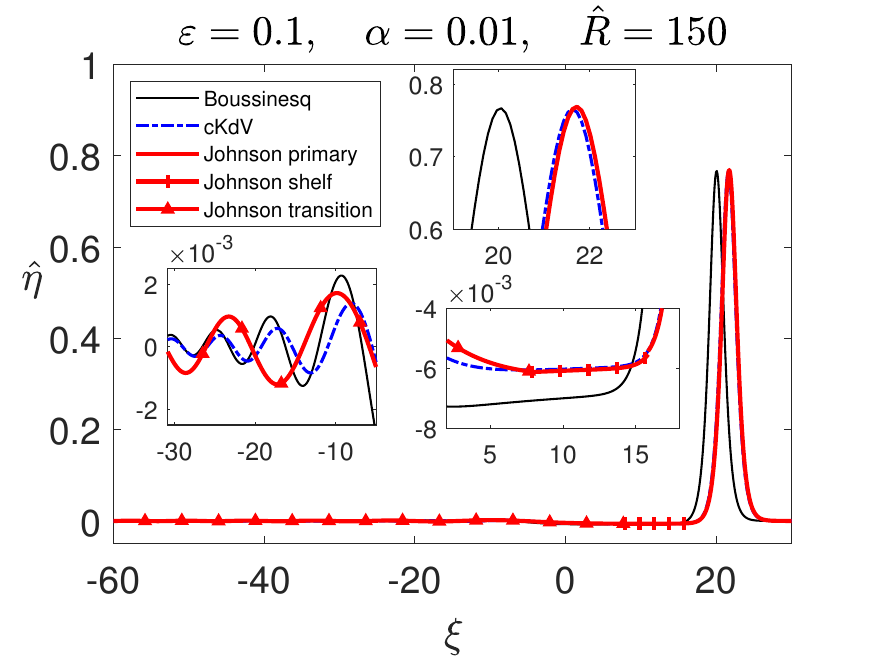}
\includegraphics[width = 8cm]{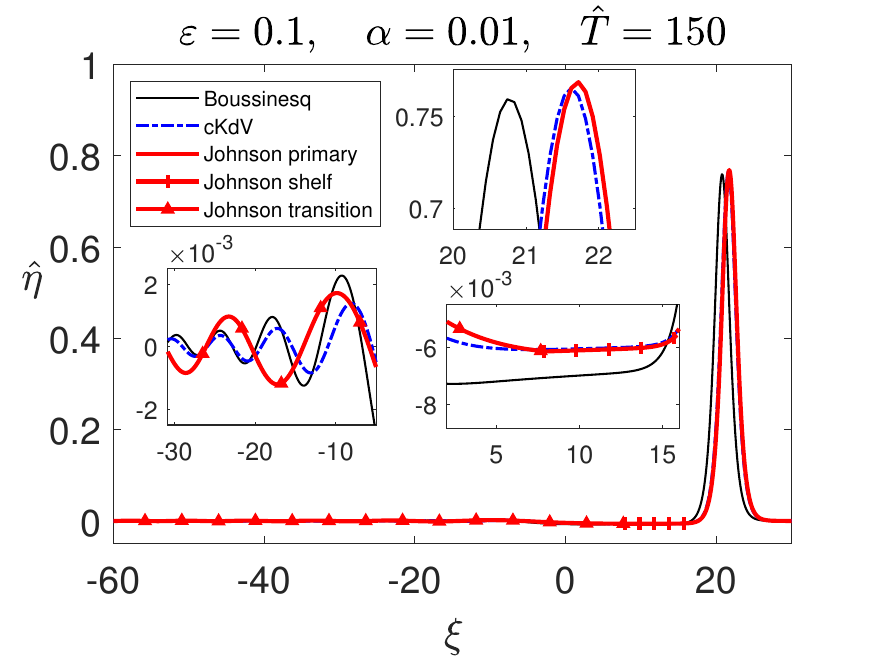}

\caption{Comparisons between axisymmetric Boussinesq, cKdV, and Johnson models for the cases $\eps \in \{ 0.01, 0.1 \}$ and $\alpha = 0.01$ over the domain $\xi \in [-90,60]$ with initial condition $\He = \sech^2 \left( \sqrt{3} \xi/2 \right)$ for all cases. Left column shows the comparisons in $(R,\xi)$-space with $R\in [\Rmin,\Rmax] = [\alpha^{-1},\alpha^{-1}+50]$ and the comparisons are made at $\Rhat \in \{ 110, 150 \}$. Right column shows these comparisons in $(T,\xi)$-space with $T \in [\Rmin,\Rmax]$ (which implies $\tau \in [\eps^{-1}\alpha^{-1},\eps^{-1}\Rmax]$). The comparisons are also made at $\That \in \{ 110, 150 \}$. Physically, the initial data for the axisymmetric Boussinesq model is placed at distance $r_0 = 10000$ away from the origin in the top comparisons, or $r_0 = 1000$ in the bottom comparisons.}


\label{Johnson_ckdv_bouss_compare}
\end{figure}

Figure \ref{Johnson_ckdv_bouss_compare} shows a comparison between Johnson's asymptotics viewed as a slow radius asymptotics (left column) vs a slow time asymptotics (right column). We see that the agreement is better in the second case. This is understandable because the axisymmetric Boussinesq problem has been solved as a time evolution problem. Therefore, in the second case the initial conditions for the cKdV equation and the Boussinesq system are the same. In the first case, the initial conditions agree only approximately. \\

\begin{figure}
\centering
\includegraphics[width = 8cm]{eps0_01_alpha0_01_JohnsonBoussinesqcKdV_T110}
\includegraphics[width = 8cm]{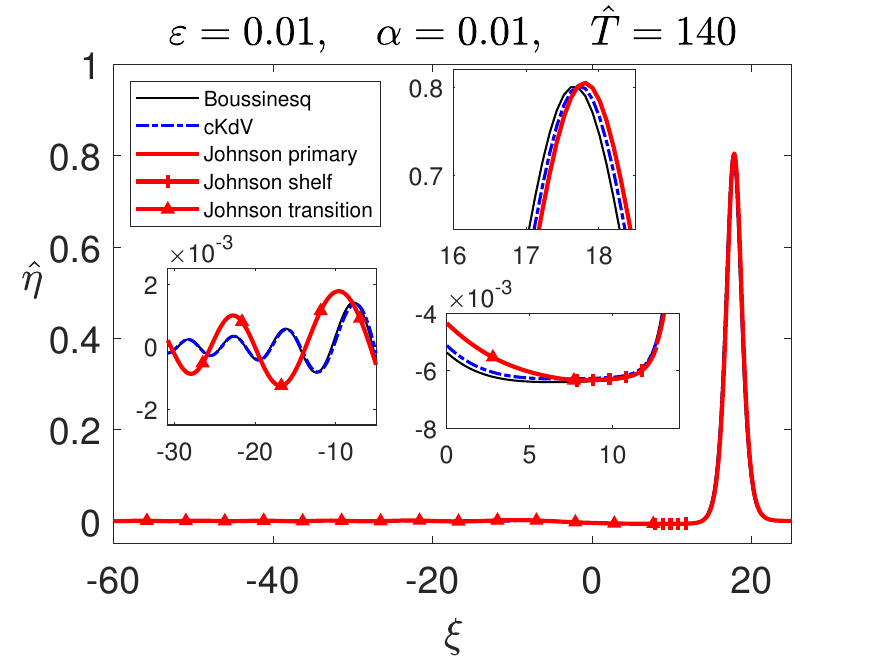}

\includegraphics[width = 8cm]{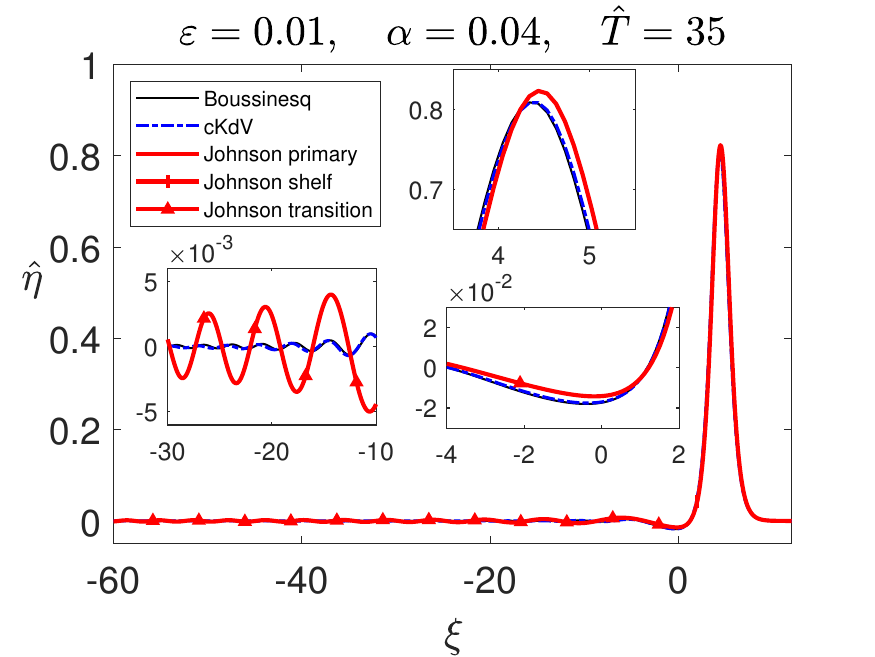}
\includegraphics[width = 8cm]{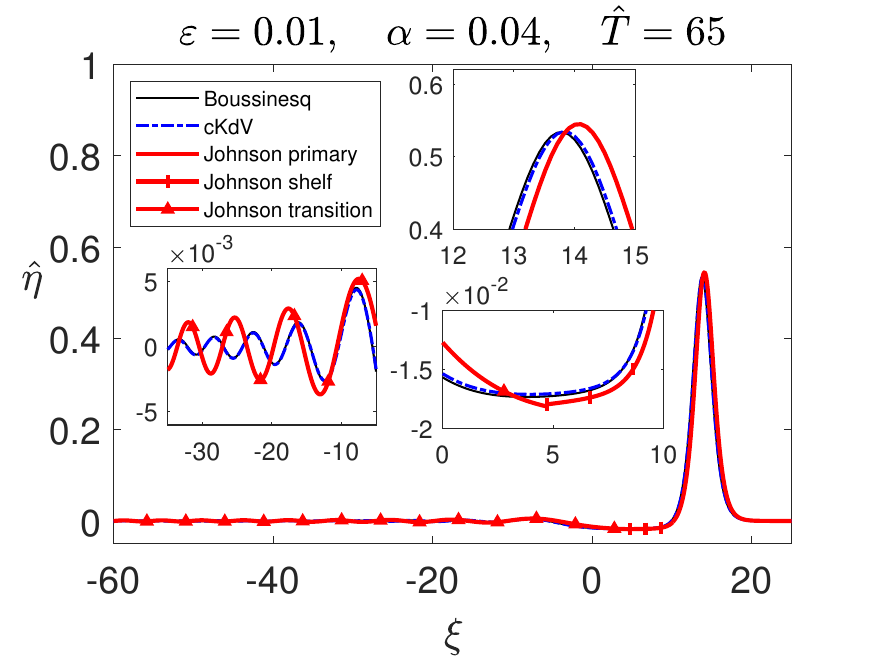}

\caption{ Comparisons between axisymmetric Boussinesq, cKdV, and Johnson models for fixed $\eps = 0.01$ and $\alpha \in \{ 0.01, 0.04 \}$ over the region $(\xi, T) \in [-90,60] \times [\alpha^{-1},\alpha^{-1} + 40]$ with initial condition $\He = \sech^2 \left( \sqrt{3} \xi/2 \right)$ in all cases. The axisymmetric Boussinesq model is evaluated over $\tau \in [\eps^{-1} \alpha^{-1}, \eps^{-1}(\alpha^{-1} + 40)]$ with initial data is placed at distance $r_0 = 10000$ away from the origin in the top comparisons, or at $r_0 = 2500$ in the bottom comparisons.}

\label{Johnson_ckdv_bouss_compare1}
\end{figure}

There are two small parameters in this problem formulation; $\alpha$ (which drives agreement between cKdV and Johnson's far-field asymptotics) and $\eps$ (which drives agreement between the weakly-nonlinear cKdV model and original Boussinesq system). 
Overall, Johnson's asymptotics gives good agreement only when both of these parameters are small enough, i.e. the waves are weakly-nonlinear, and the initial condition is placed sufficiently far away from the origin. As shown in  Figure \ref{Johnson_ckdv_bouss_compare1} for $\eps = 0.01$ and the two values of $\alpha \in \{ 0.01, 0.04 \}$, Johnson's solution starts to deviate  from other solutions at an earlier time for a greater value of $\alpha$. 
 In particular, the oscillatory transition region approximation is valid only at large distances away from the initial position of the pulse. This is understandable, because the approximation is based only on the known level of the shelf, and not on its slope. The long-time asymptotics of the emerging undular bore does not depend on the slope \cite{B}, and hence the approximation works sufficiently far away from the initial position of the pulse.
When $\eps$ or the wave amplitude is increased, it can be observed that the cKdV solution and, therefore, its asymptotic solution become invalid at a much earlier time in comparison with the Boussinesq equations.\\

\begin{figure}
\centering
\includegraphics[width = 8cm]{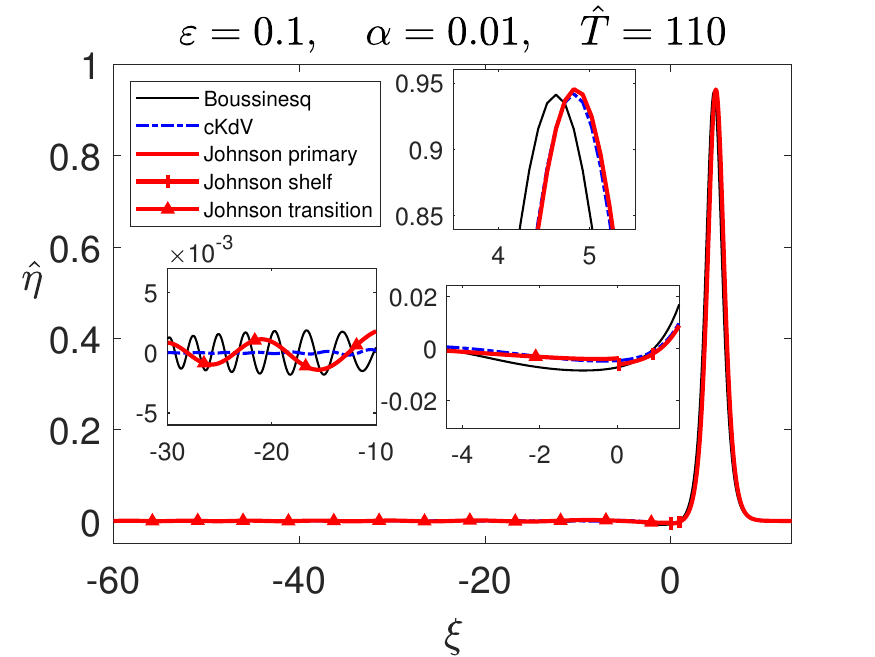}
\includegraphics[width = 8cm]{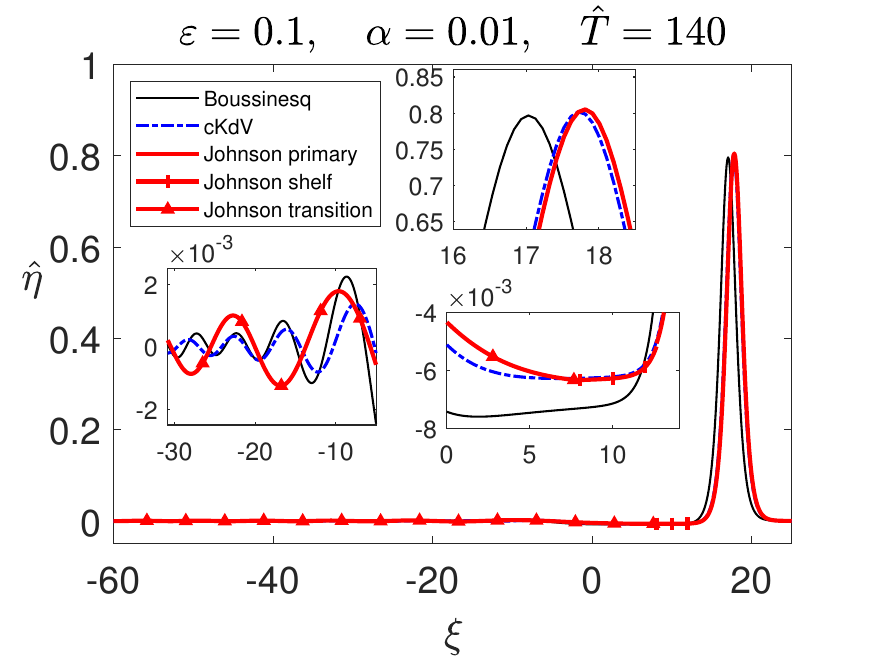}

\includegraphics[width = 8cm]{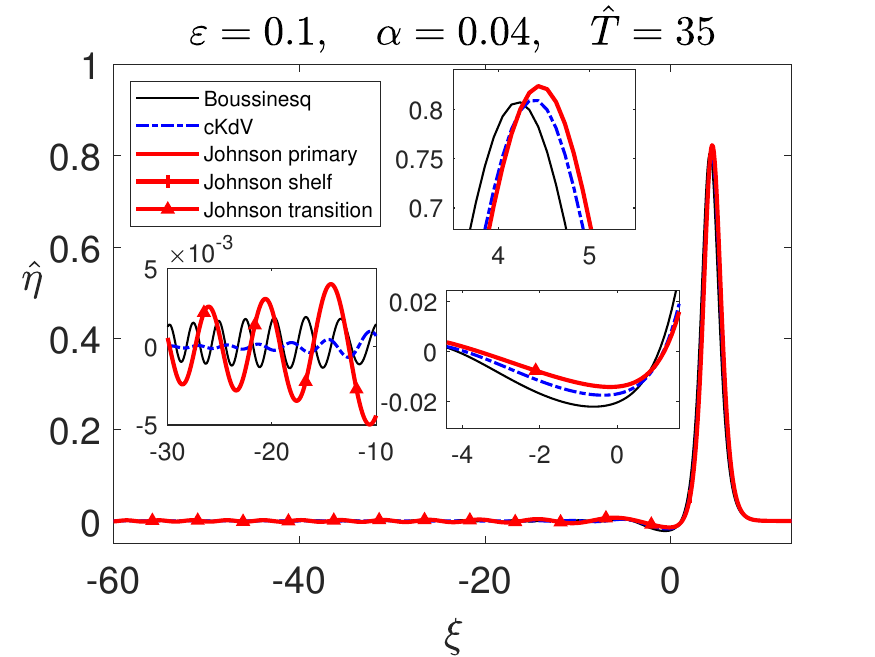}
\includegraphics[width = 8cm]{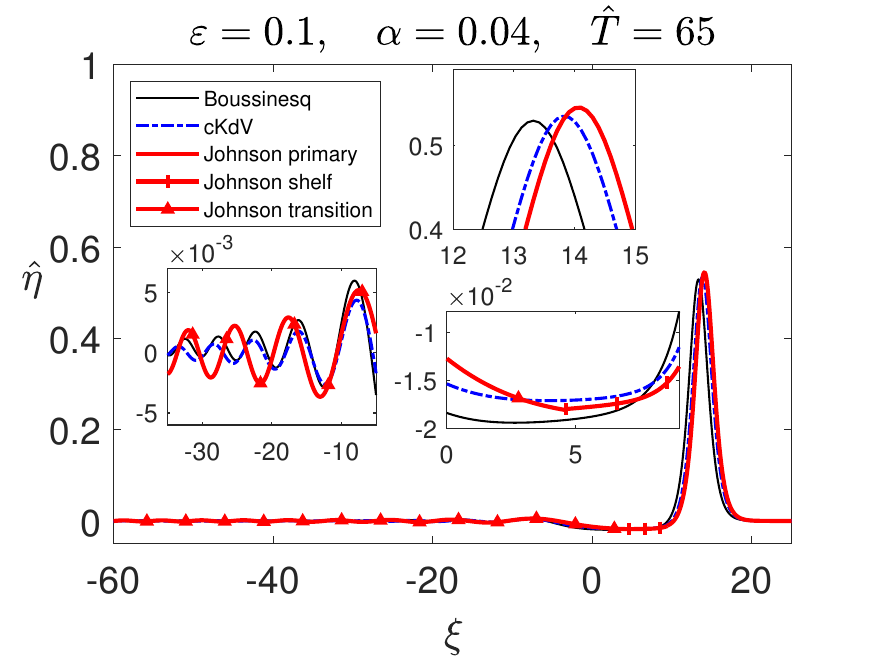}

\caption{Comparisons between axisymmetric Boussinesq, cKdV, and Johnson models for fixed $\eps = 0.1$ and $\alpha \in \{ 0.01, 0.04 \}$ over the region $(\xi, T) \in [-90,60] \times [\alpha^{-1},\alpha^{-1} + 40]$ with initial condition $\He = \sech^2 \left( \sqrt{3} \xi/2 \right)$ in all cases. The axisymmetric Boussinesq model is evaluated over $\tau \in [\eps^{-1} \alpha^{-1}, \eps^{-1}(\alpha^{-1} + 40)]$ with initial data is placed at distance $r_0 = 1000$ away from the origin in the top comparisons, or at $r_0 = 250$ in the bottom comparisons.}

\label{Johnson_ckdv_bouss_compare2}
\end{figure}

Finally, Figure \ref{Johnson_ckdv_bouss_compare2} shows a comparison between the Boussinesq, cKdV and Johnson's solutions for $\eps = 0.1$ and $\alpha \in \{ 0.01, 0.04 \}$. 
The agreement between the solutions of the axisymmetric Boussinesq system and the cKdV equation is now much worse than in Figure \ref{Johnson_ckdv_bouss_compare1}. Johnson's asymptotics continues to be a rather good approximation to the solution of the cKdV equation for the primary wave and the shelf, but there is a significant phase shift and amplitude difference compared to the solution of the full Boussinesq system. As before, the behaviour of the oscillatory tail region is captured well only at very large values of $T$ due to the nature of this approximation. 
\\

\section{Numerical simulations: cKdV vs ecKdV models}
\label{sec:Num2}

 \begin{figure}[h]
\centering
\includegraphics[width = 8cm]{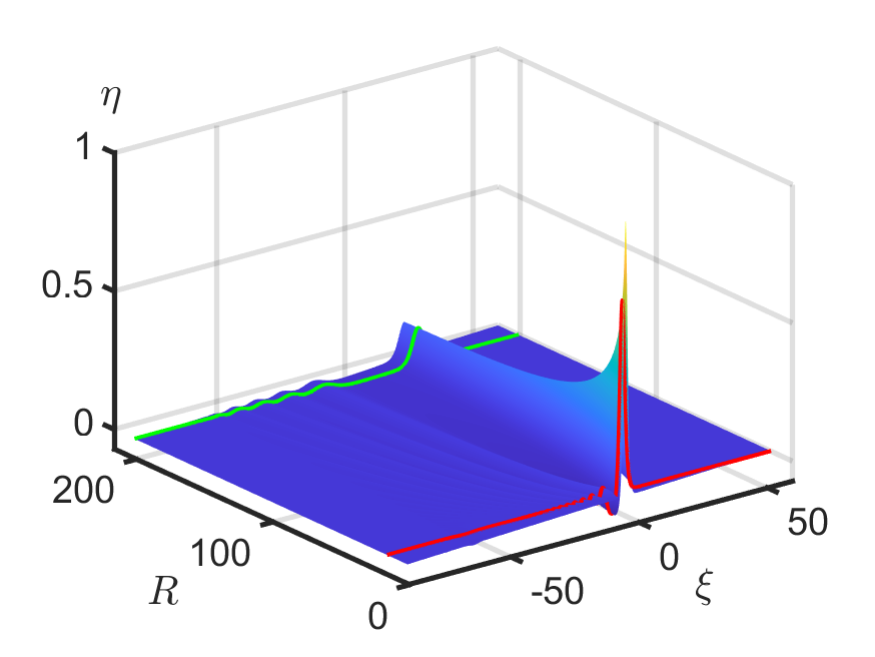}
\includegraphics[width = 8cm]{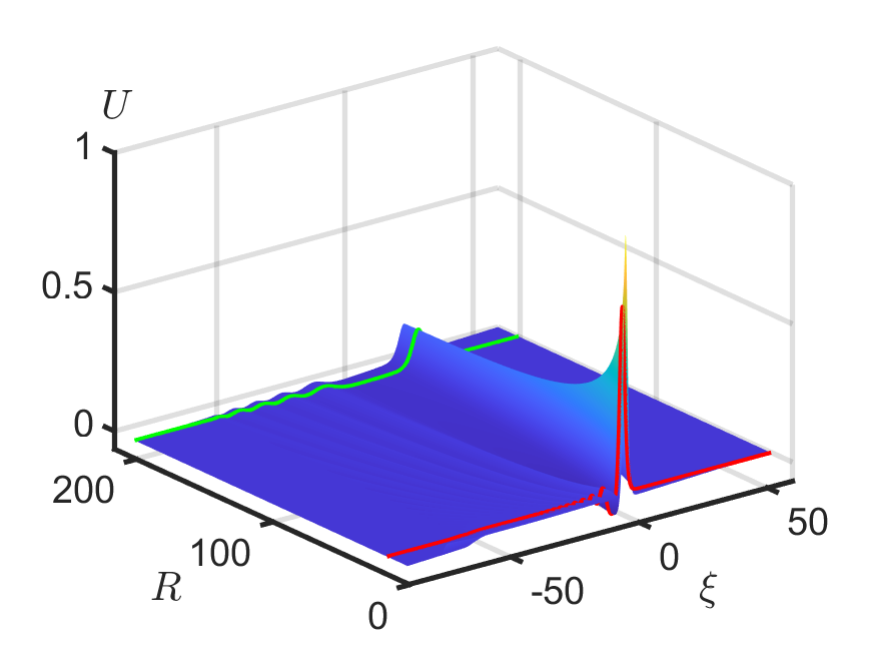}
\caption{Boussinesq $\eta,U$ surfaces in the $(R,\xi)$-space with $\eps = 0.1$, $\tau \in [\taumin,\taumax] = [100,2100]$, $\xi \in [-90,60]$ and initial condition $\eta_0 = \sech^2(\xi)$. The solid red line denotes the initial condition for cKdV and ecKdV models taken at $\Rmin = \eps (\ximax + \taumin) = 16$, whereas the solid green line denotes the maximum distance from the origin where the comparison of the models can be made. This is calculated as $\Rmax = \eps(\ximin + \taumax) = 201$. The initial condition for Boussinesq system has been placed at the physical distance $r_0 = \taumin = 100$ away from the origin, with cKdV and ecKdV models being used once the wave passes the distance mark of $r=160$.}

\label{eps0_1_xiR_surface}
\end{figure}

In this section we perform a detailed comparison of the accuracy of the cKdV and ecKdV models when they are used to speed up the solution of an initial-value problem for the axisymmetric Boussinesq system, as detailed in Section \ref{sec:Num}.\\

We first investigate the case of $\eps = 0.1$ and observe the evolution of $\eta,U$ for long times $\tau$. There is very little observable difference between $\eta,U$ at the later stages of evolution, and we see that a long shelf behind the decaying wave-front develops before transitioning into the oscillatory region. Once the model is fully evaluated, we can then plot a surface of all $\eta$ profiles in the $(\xi,\tau)$-space.  In fact, using the relation 
\begin{equation}
R = \eps \tilde{r} = \eps (\tau + \xi),
\end{equation}
 we plot this surface in the $(\xi,R)$-space instead as shown in Figure \ref{eps0_1_xiR_surface}. We can extract a cut of this surface along $R=\Rmin$ (the solid red line) using cubic spline interpolation before using this as initial data for cKdV (\ref{ckdv}) and ecKdV (\ref{eckdv}) models. We can then numerically integrate these two models up to $R = \Rhat$ before making a comparison between all three at this point. \\

\begin{figure}[h]
\centering
\includegraphics[width = 8cm]{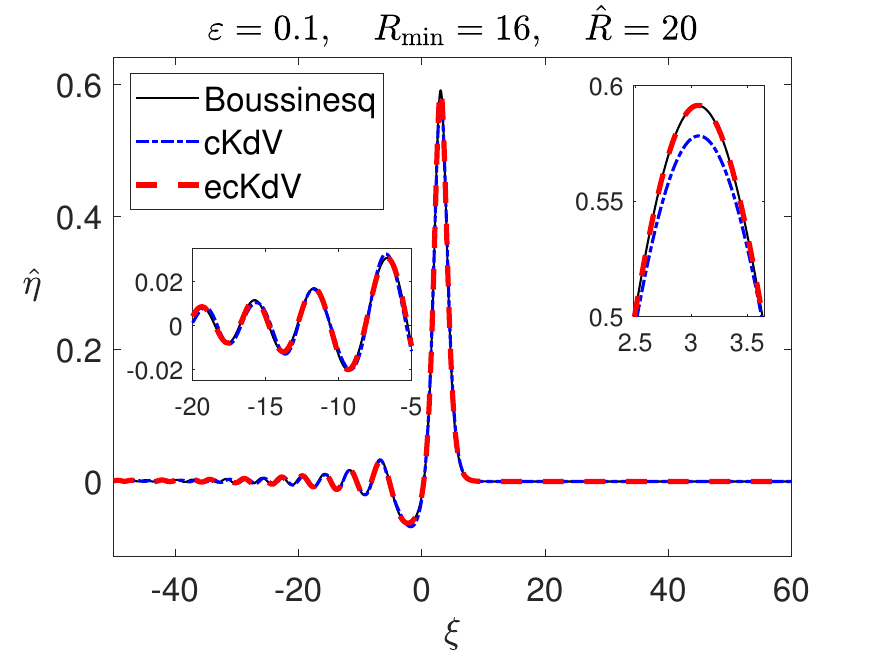}
\includegraphics[width = 8cm]{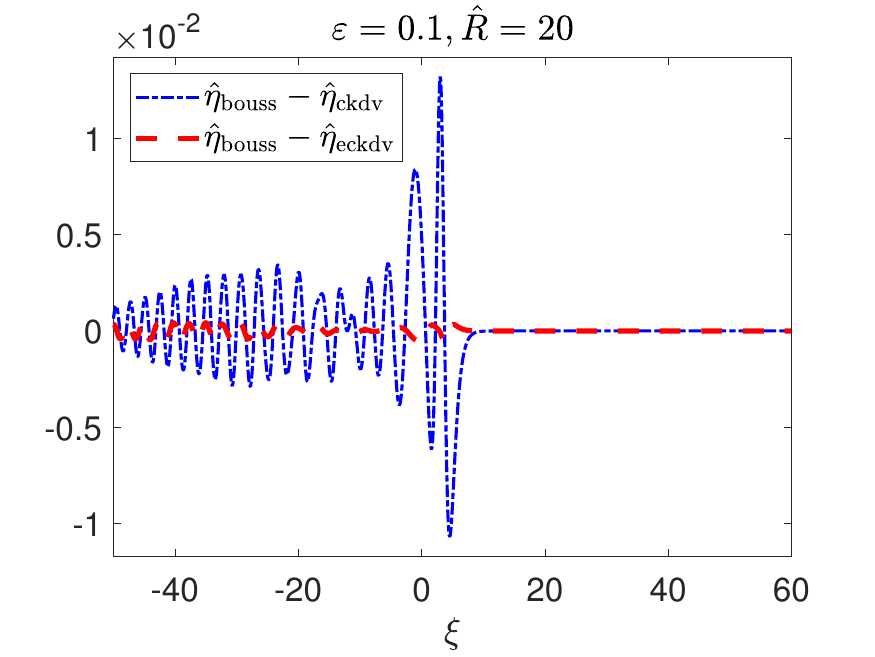}
\includegraphics[width = 8cm]{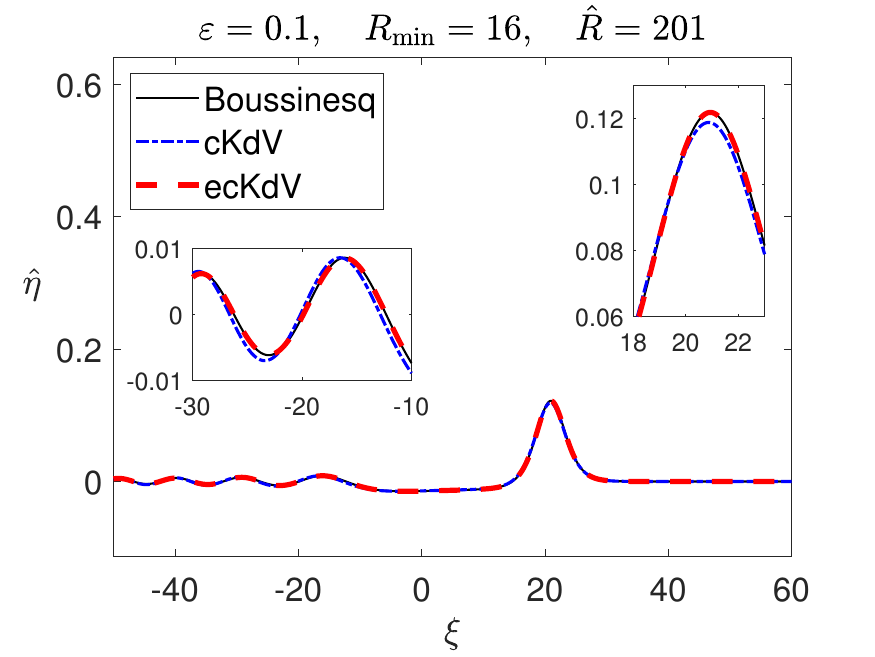}
\includegraphics[width = 8cm]{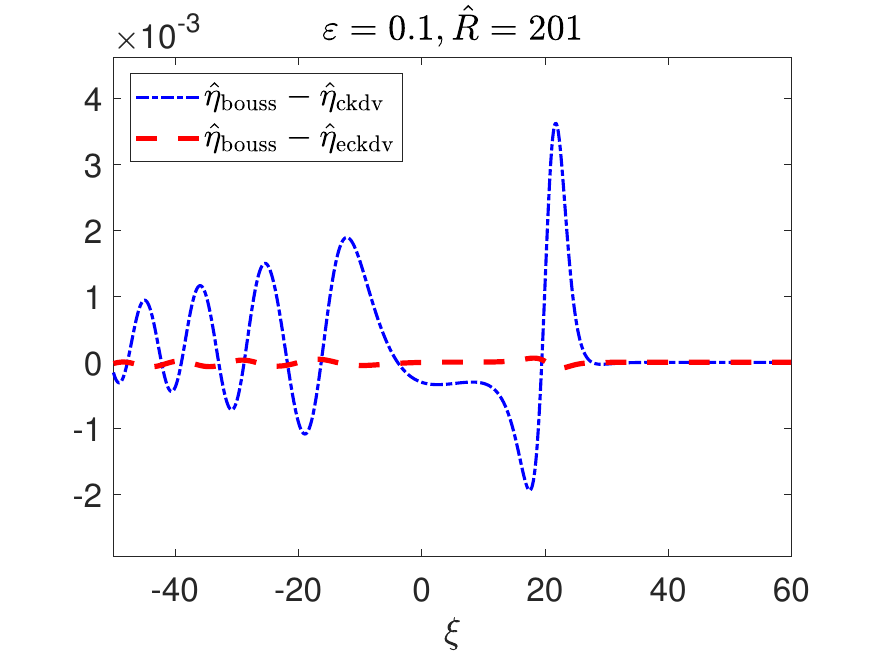}
\caption{Comparison between axisymmetric Boussinesq, cKdV, and ecKdV models for the case $\eps = 0.1$ with Boussinesq having initial data $\eta_0 = \sech^2 \xi$ and being integrated over the region $(\xi, \tau) \in [-90,60] \times [100,2100]$. This implies placement of cKdV-type initial data at $R = \Rmin = 16$ with comparisons taking place at $R = \Rhat \in \{20,201\}$. Physically, the Boussinesq initial data here is placed at distance $r_0 = 100$ from the origin, and cKdV-type models are employed at distance $r=160$.}


\label{eps0_1_modelcomparisonsETA}
\end{figure}

\begin{figure}[h]
\centering
\includegraphics[width = 8cm]{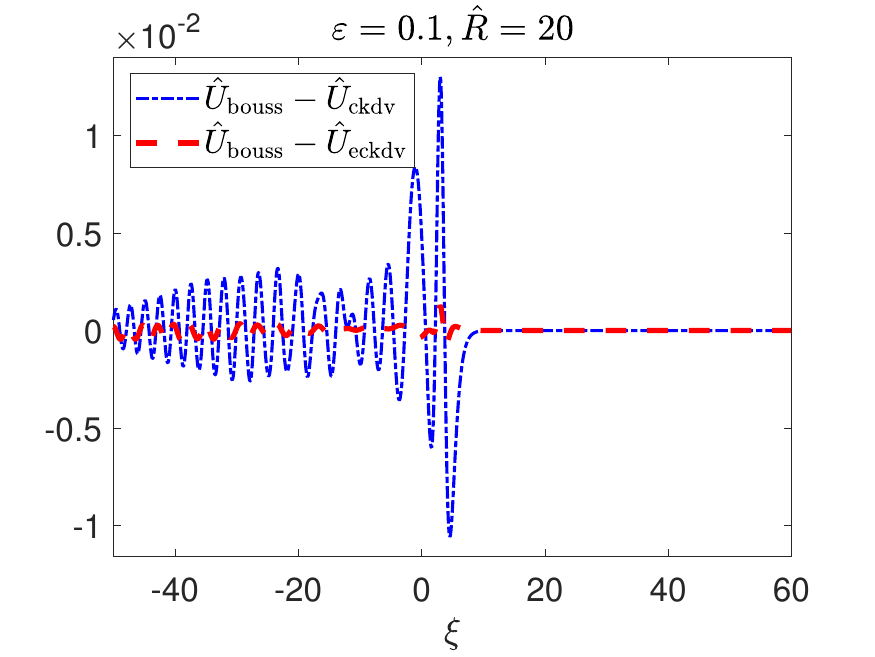}
\includegraphics[width = 8cm]{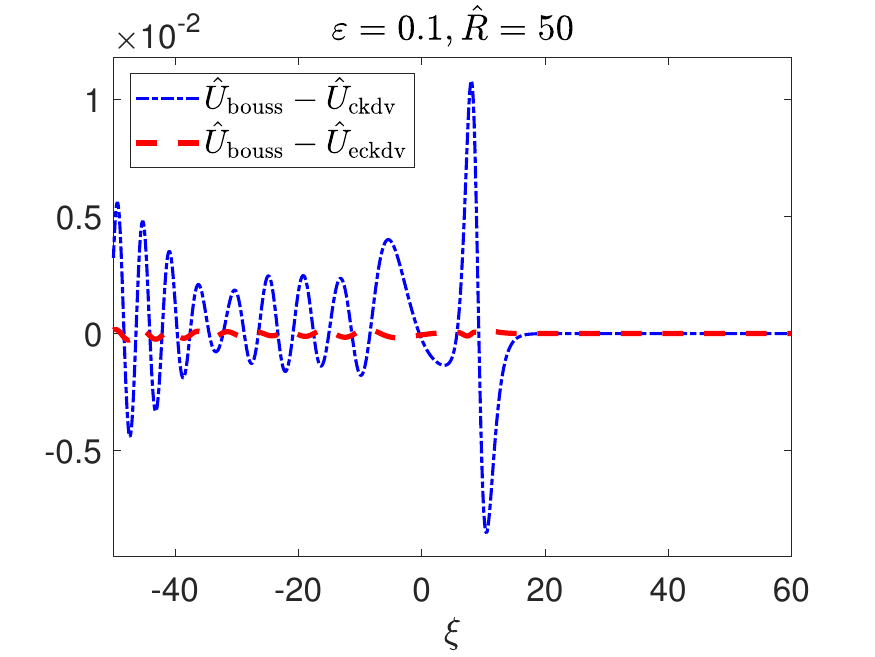}
\includegraphics[width = 8cm]{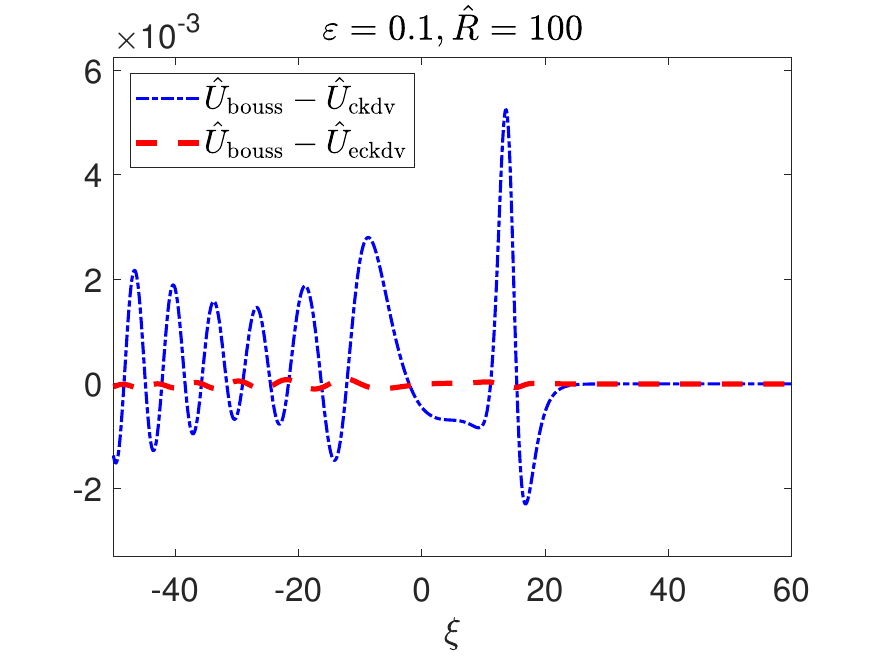}
\includegraphics[width = 8cm]{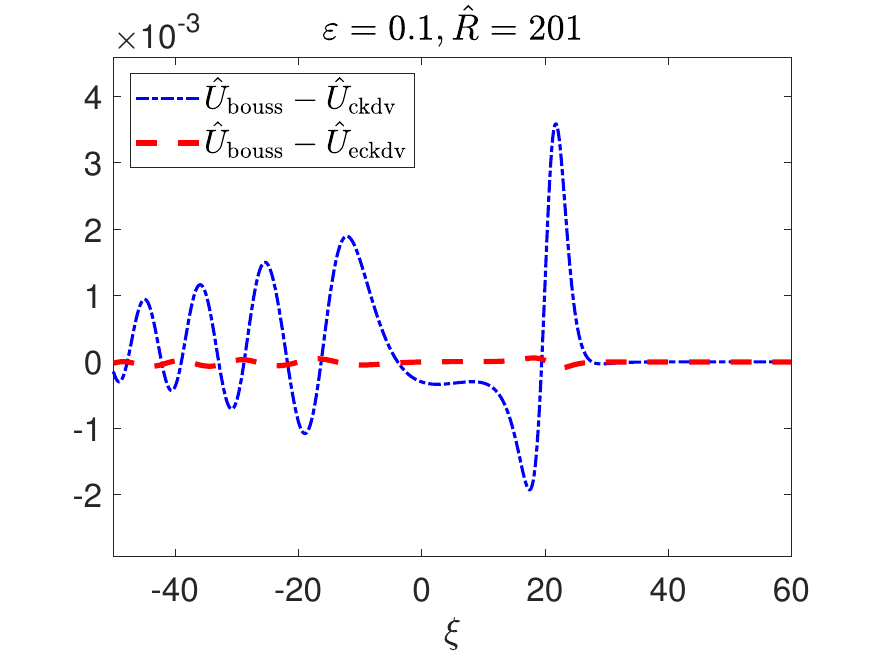}
\caption{Error plots for $\HU$ between axisymmetric Boussinesq, cKdV, and ecKdV models for the case $\eps = 0.1$ with Boussinesq system having initial condition $\eta_0 = \sech^2 \xi$ and being integrated over the region $(\xi, \tau) \in [-90,60] \times [100,2100]$. This implies placement of cKdV-type initial data at $R = \Rmin = 16$ with comparisons taking place at $R = \Rhat \in \{20,50, 100, 201\}$. Here we compare exact $\HU$ from Boussinesq system against $\HU$ as given by the relation $\displaystyle \HU = \He + \dfrac{\eps}{2\Rhat} \int_{\xi}^{\ximax} \He \ d\xi - \eps \left( \dfrac{1}{4} \He^2 - \dfrac{1}{6} \He_{\xi\xi} \right)$ where $\He \in \{\He_{\text{ckdv}},\He_{\text{eckdv}}\}$.}

\label{eps0_1_modelcomparisonsU}
\end{figure}
 
The detailed comparisons between all three models have been made for $\Rhat \in \{ 20, 50, 100, 201 \},$ and Figure \ref{eps0_1_modelcomparisonsETA} shows the comparisons at $\Rhat \in \{ 20, 201 \}.$ A key observation we could make was that the ecKdV equation was much closer to the solution of the full axisymmetric 2D Boussinesq model than the solution of the cKdV equation for all sampled $R$ values. The three key regions of the wave, i.e. primary wave, shelf and oscillatory region, are all well described by the ecKdV equation. The cKdV equation underestimates  the amplitude of the solution within the primary wave region, which is a consistent discrepancy. On the other hand, the ecKdV model retains no visual difference from the axisymmetric 2D Boussinesq system over the entire computational domain. 
Moreover, the difference between the cKdV and ecKdV $\He$ solutions and the corresponding solution obtained using direct numerical simulations of the axisymmetric Boussinesq system across the $\xi$ domain can be tracked via the right-hand panels of Figure \ref{eps0_1_modelcomparisonsETA}, which show that the ecKdV models give a much smaller error, and this was true for all the sampled cases. The difference between the cKdV and ecKdV models is rather significant as is illustrated for $\Rhat = 20$ and $\Rhat = 201$. 
Naturally, both errors reduce with the distance. 
Similar comparisons for the radial velocity $\HU$ are shown in Figure \ref{eps0_1_modelcomparisonsU} for $\Rhat \in \{ 20, 50, 100, 201 \},$  and the errors are again much smaller when we use the ecKdV model
\\
 
 \begin{figure}[h]
\centering
\includegraphics[width = 8cm]{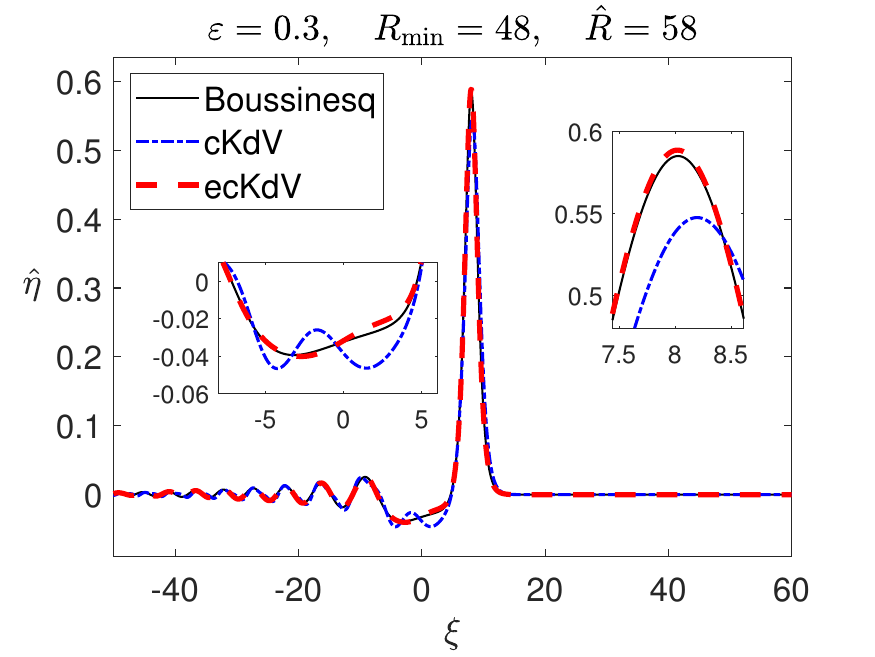}
\includegraphics[width = 8cm]{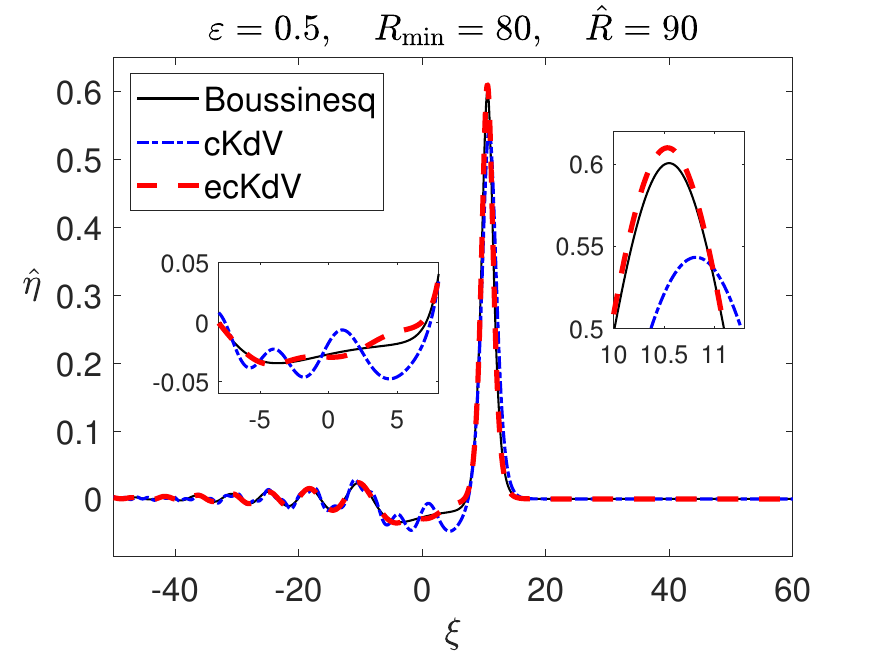}

\caption{Comparison between axisymmetric Boussinesq, cKdV, and ecKdV models for the cases $\eps \in \{ 0.3, 0.5\}$ with Boussinesq system having initial condition $\eta_0 = \sech^2 \xi$ and being integrated over the region $(\xi, \tau) \in [-90,60] \times [100,500]$. This implies placement of cKdV-type initial data at $R = \Rmin = 48$ (when $\eps = 0.3$) or $R = \Rmin = 80$ (when $\eps = 0.5$) with comparisons taking place at $R=\Rhat=\Rmin+10$. The initial condition for Boussinesq system has been placed at the physical distance $r_0 = 100$ away from the origin, and cKdV-type models are employed at distance $r=\eps^{-1}\Rmin=160$ in both cases.}


\label{eps0_5_comparison}
\end{figure}

\begin{figure}
\centering
\includegraphics[width = 8cm]{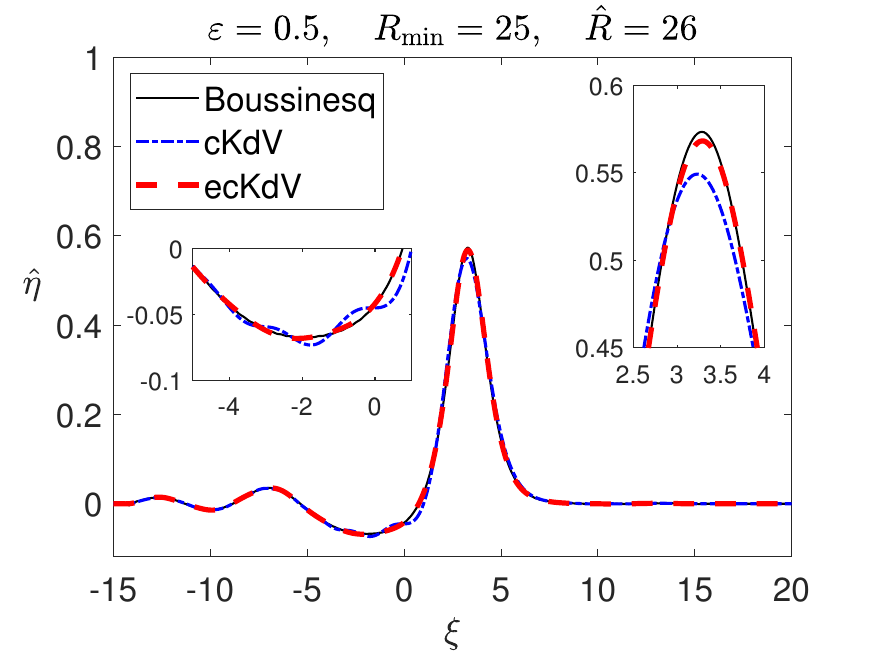}
\includegraphics[width = 8cm]{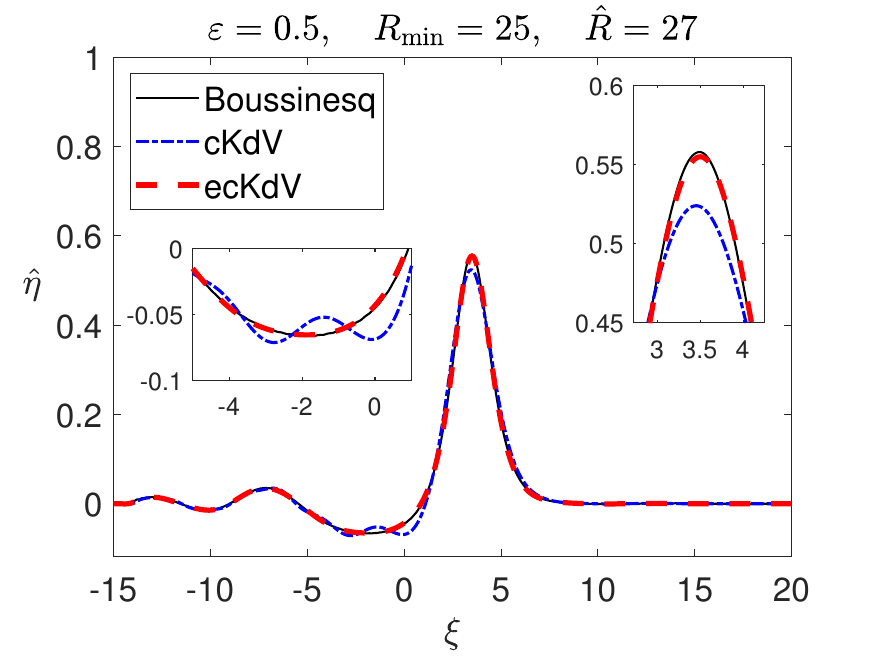}

\includegraphics[width = 8cm]{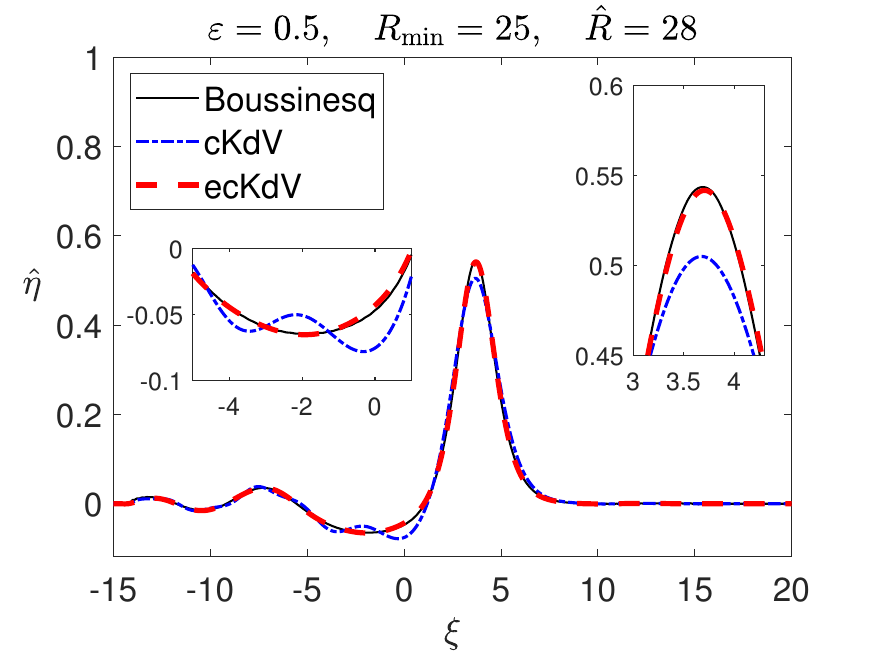}
\includegraphics[width = 8cm]{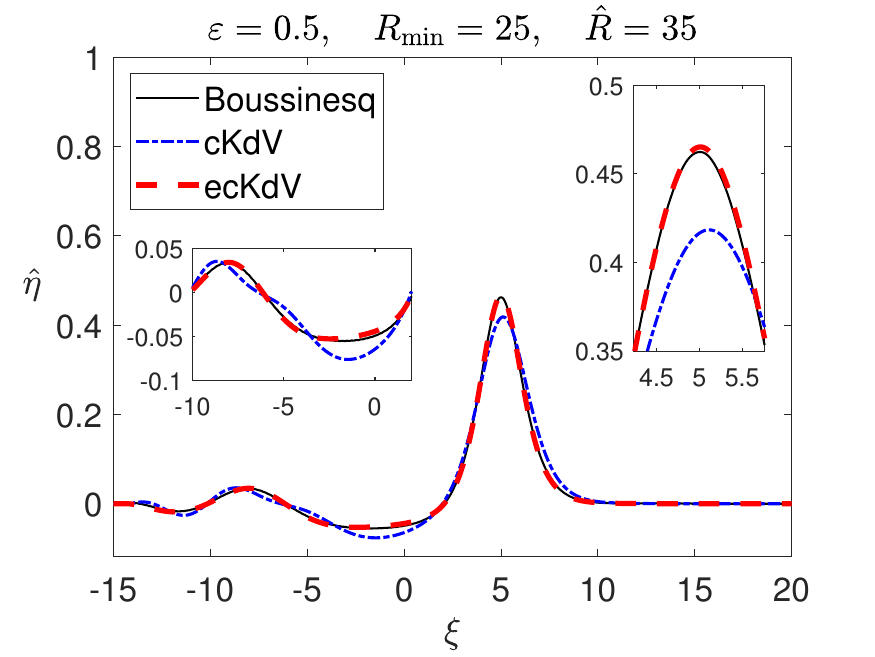}

\includegraphics[width = 8cm]{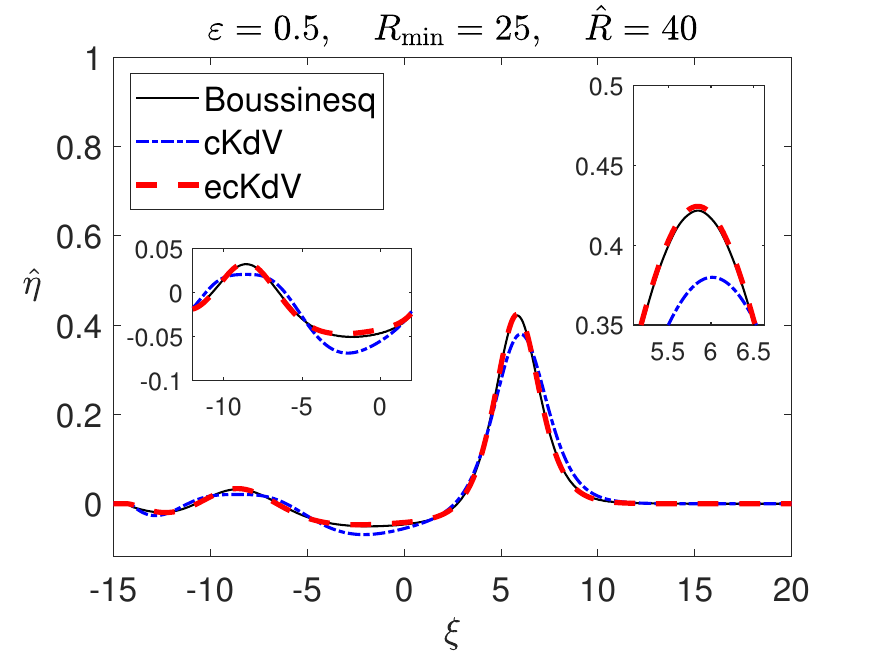}
\includegraphics[width = 8cm]{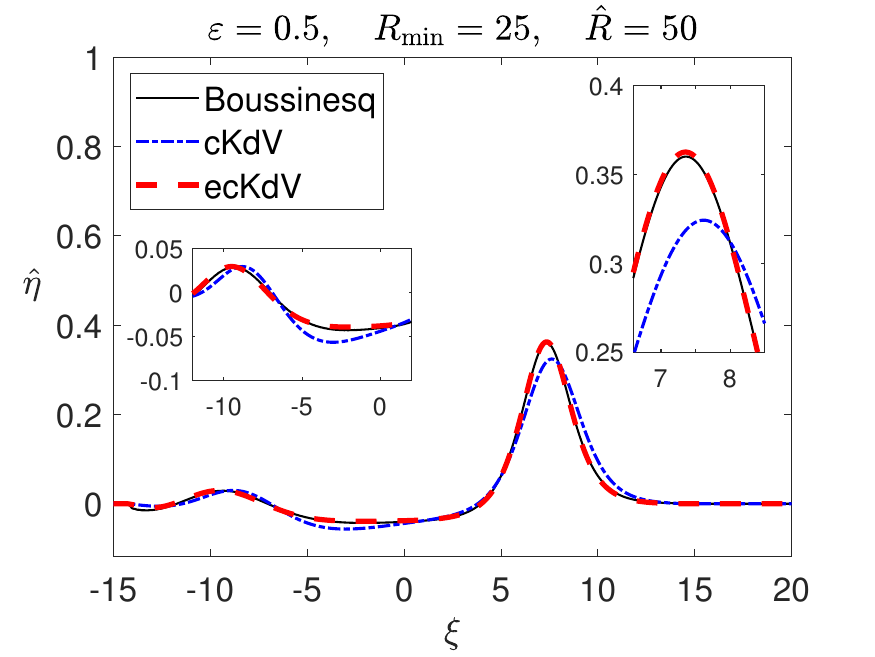}

\caption{Simulations of the three models with $\eps = 0.5$ on the adjusted Boussinesq region $(\xi,  \tau) \in [-24,25] \times [25,154]$ to allow comparisons to be made at significantly smaller $\hat{R} \geq 25$ values. The initial condition for Boussinesq system remains as $\eta_0 = \sech^2 \xi$, with suppression functions $F,s$ requiring adjusting $\kappa_F = \kappa_s = 10$ and $\xi_{\text{span}}$ factor increasing from $1/10$ to $2/10$ which allows waves near the boundaries of this shorter domain to be suppressed in good time. Physically, the Boussinesq initial data here is placed at distance $r_0 = 25$ from the origin, and cKdV-type models are employed at distance $r=\eps^{-1}\Rmin=50$.}

\label{eps0_5_comparison_smallR}
\end{figure}

In Figures \ref{eps0_5_comparison} and  \ref{eps0_5_comparison_smallR} we show a comparison between the Boussinesq, cKdV and ecKdV models for the cases when the waves have not small but moderate amplitude ($\eps \sim 0.3 - 0.5$). Figure \ref{eps0_5_comparison} shows the results for $\eta$ related to the comparison of solutions of the three models for the case when the Boussinesq run has been initiated at $r_0 = 100$ with $\eps \in \{ 0.3, 0.5 \}$. Such relatively large values of the amplitude parameter are clearly outside of the range of validity of the cKdV equation, which develops large spurious oscillations in the shelf region already for $\varepsilon = 0.3$.  The differences are even more significant for $\eps = 0.5$.  The ecKdV equation captures the solution rather well in both cases, and visibly better than the cKdV equation. Next, Figure \ref{eps0_5_comparison_smallR} shows a similar and more detailed comparisons for simulations related to the case when the Boussinesq run was initiated much closer to the origin, at $r_0 = 25$ with $\eps = 0.5$. In these runs $\Rmin = 25$ and  comparisons are shown up to $\hat R = 50$. The ecKdV model stays on top of the Boussinesq solution at all times, while the cKdV solution has visible discrepancies. We do not show the error plots for these large values of $\eps$ since the differences between the solutions are clearly visible in the main plots. \\
  
 
\begin{figure}[h]
\centering
\includegraphics[width = 8cm]{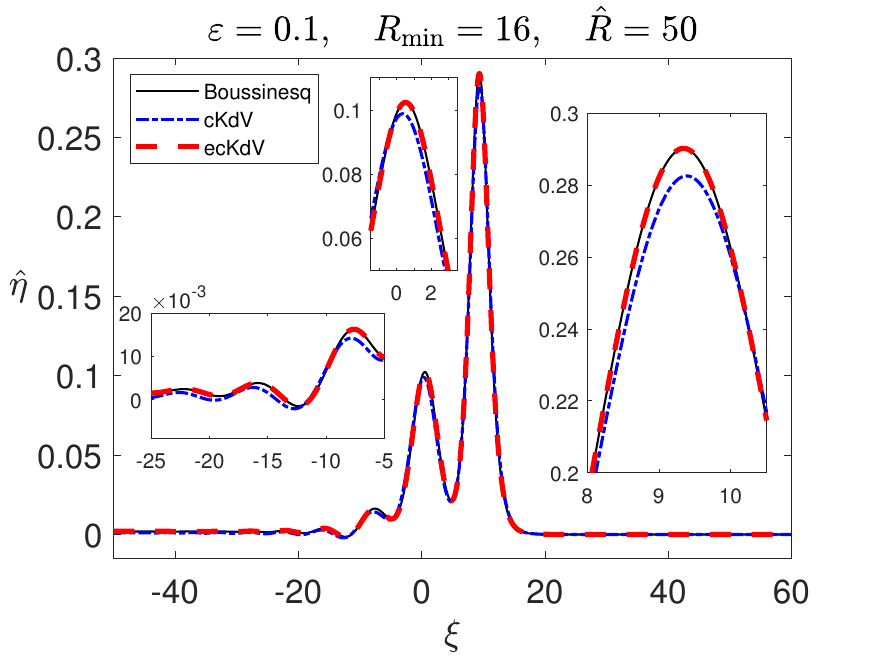}
\includegraphics[width = 8cm]{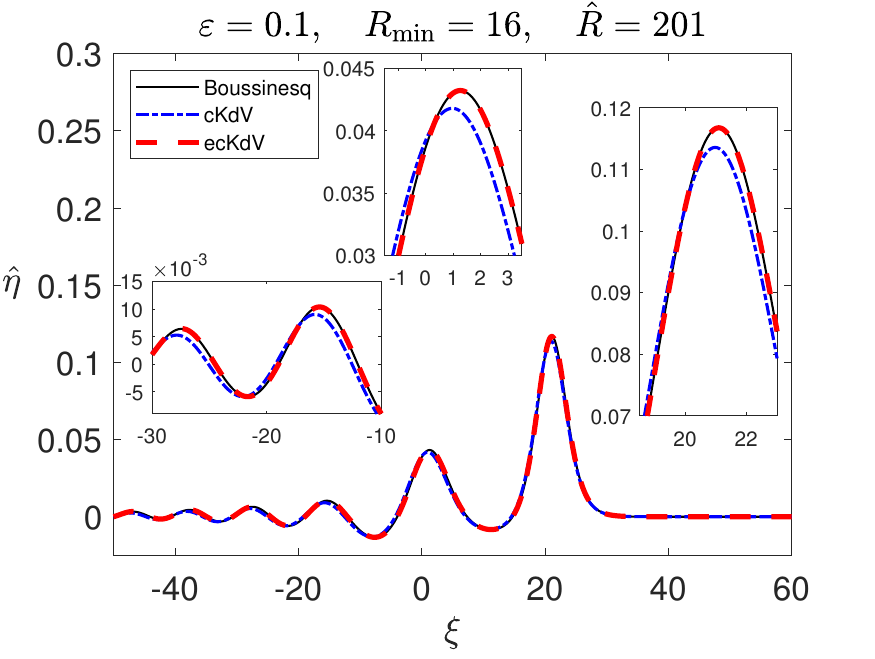}
\caption{Comparison between axisymmetric Boussinesq, cKdV, and ecKdV models for the case $\eps = 0.1$ with Boussinesq system having initial condition $\eta_0 = 0.5 \sech^2 (0.25 \xi)$ and being integrated over the region $(\xi, \tau) \in [-90,60] \times [100,2100]$. This implies placement of cKdV-type initial data at $R = \Rmin = 16$ with comparisons taking place at $R=\Rhat$. In particular, this choice of initial data $\eta_0$ generates fission within the simulation, and comparisons are made at $\Rhat \in \{50,201\}$. The initial condition for Boussinesq system has been placed at the physical distance $r_0 = 100$ away from the origin.}

\label{eps0_1_fission}
\end{figure}

The next natural question to ask is what happens if we vary the parameters of the initial profile for the Boussinesq system. By changing from $(A,\lambda) = (1,1)$ to $(A,\lambda) = (0.5,0.25)$, we can see the results in Figure \ref{eps0_1_fission}. For this wide initial profile we observe fission, i.e. the primary wave splits into two; with one fast pulse ahead, and another slower pulse behind. This is a complicated process which cKdV-type models should replicate. We see that at the smallest value $\Rhat = 50$ the cKdV model is already having significant problems trying to match the Boussinesq system. Both pulses are underestimated in amplitude and mismatched in phase. The same can be said about the oscillatory region, and this discrepancy is present for all sampled $R$ values. On the other hand, the ecKdV equation captures the solution of the Boussinesq model very well once again with no noticeable difference.
\\

\begin{figure}[h]
\centering
\includegraphics[width = 8cm]{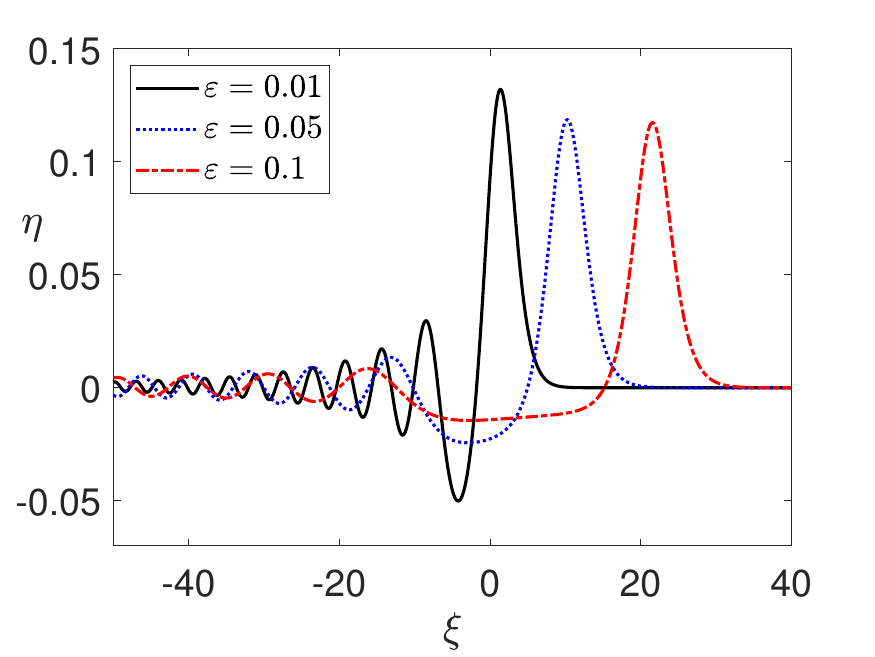}
\includegraphics[width = 8cm]{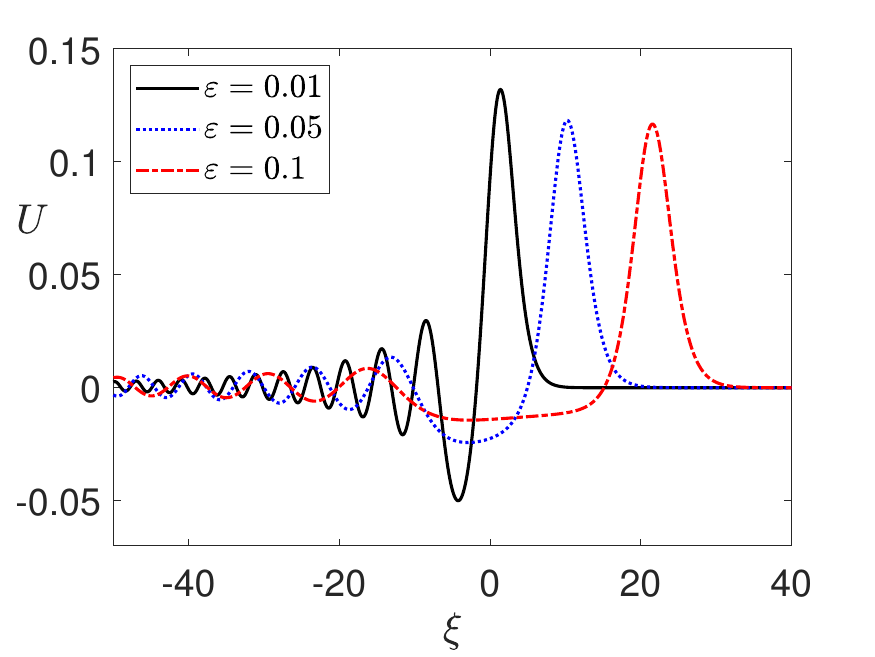}

\caption{The final axisymmetric Boussinesq profiles at $\tau = 2100$ for different $\eps$ cases. The initial condition is $\eta_0 = \sech^2(\xi)$ across all cases. Physically, all initial profiles were placed at distance $r_0 = 100$ away from the origin with cKdV-type models being employed at $r = 160$.}

\label{final_profiles}
\end{figure}

In our studies, we experimented with different values of the small-amplitude parameter. The main features of  the solution mainly remain the same as shown in Figure \ref{final_profiles}, i.e. the primary wave is followed by a shelf and an oscillatory region.  The behaviour of the primary wave is qualitatively similar to that described by the Inverse Scattering Transform for the Korteweg - de Vries equation \cite{GGKM}, i.e. depending on the amplitude and the width of the initial condition it might fission into a number of solitons and radiation, which then evolve under the mutual action of nonlinear, dispersive and cylindrical divergence terms. We note that, in the cases considered by us, the discrete spectrum of the Schr\"odinger equation associated with the relevant KdV equation (see \cite{GGKM}) gave rather good predictions for the initial fission of the primary wave.\\

\begin{figure}[h]
\centering
\includegraphics[width = 8cm]{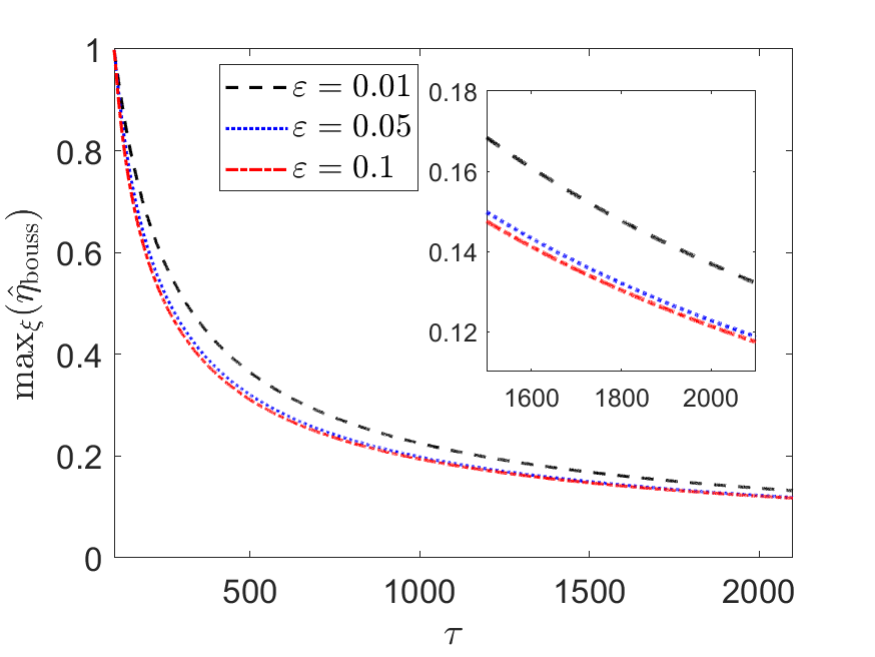}
\includegraphics[width = 8cm]{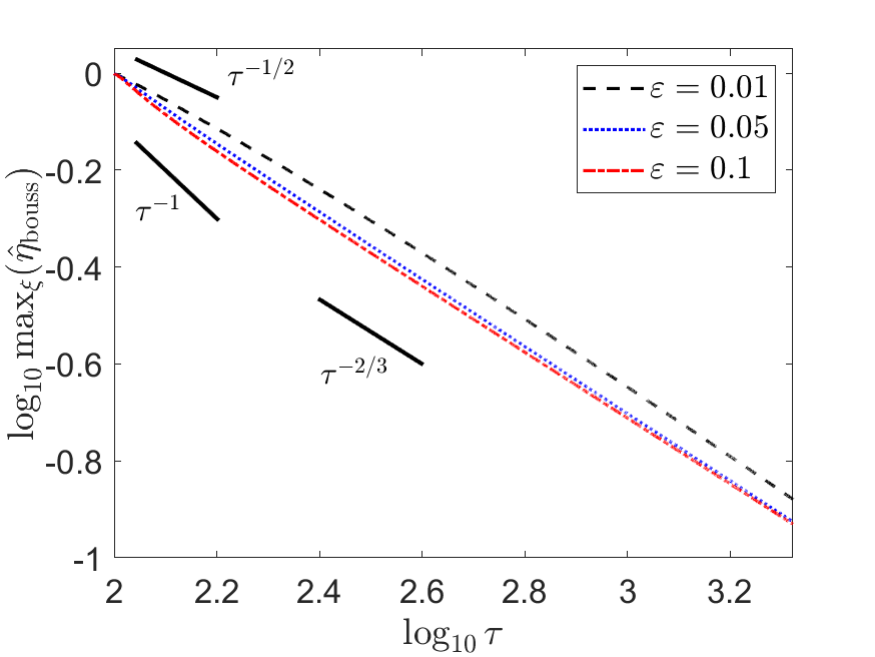}
\caption{The amplitude of $\eta$ pulse in axisymmetric Boussinesq system as $\tau \in [100,2100]$ varies for different $\eps$ cases. There are three theoretical power laws for comparison: lines with gradients $-1/2, -2/3$ and $-1$. The three initial slopes are $\{ -0.5263, -0.6316, -0.7895 \}$ in order from top to bottom. }
\label{max_eta_diff_eps}
\end{figure}

By emulating the axisymmetric 2D Boussinesq system up to $\tau = 2100$ for an initial profile with $(A,\lambda) = (1,1)$, we see that large $\eps$ leave a longer shelf behind and the wavelength of the oscillatory region is reduced. The amplitudes of the primary waves decay, which was extensively studied (e.g. \cite{S, HRS}). Our numerical results are in agreement with these studies as shown in Figure \ref{max_eta_diff_eps}. For smaller waves, the initial decay rate is closer to $\sim r^{-1/2}$, which corresponds to the dominant effect of cylindrical divergence. For larger waves of the same width, it is closer to $\sim r^{-1}$, which corresponds to the combined cylindrical divergence and dispersive effects (waves are steeper). However, at a later time, all decay rates approach the $\sim r^{-2/3}$ power law, corresponding to the leading-order balance of the nonlinear, dispersive and cylindrical divergence terms. \\

Finally, we note that when we reduce $\eps$, we observe much less difference between all three models (as one might expect based on the theoretical results). This means that for small $\eps$ values which describe small-amplitude cylindrical waves, while the ecKdV equation is still more accurate than the cKdV model, the accuracy of the cKdV model might be enough for practical applications. However, we would like to emphasise that the ecKdV model performs significantly better for larger values of  $\eps$, describing waves of moderate amplitude. \\

\section{Concluding remarks}
\label{sec:Conclusion}

In this paper, we have considered nonlinear outward-propagating concentric waves in shallow water using the 2D Boussinesq system and the reduced models. Our main aim is to investigate how good the extended cKdV equation is in comparison with
the leading-order cKdV equation in the generic situation, i.e. when all coefficients of the leading-order equation are $O(1)$.\\

 In order to do that, we have chosen the axisymmetric Boussinesq system previously modelled numerically for inward-propagating waves
 by Chwang and Wu \cite{CW} as our primitive equations. A pseudospectral method has been used to 
 accurately solve the system for a relatively long period of time.
 A filter has been applied to the localised initial conditions and computed numerical solutions  in order to suppress any disturbances near the boundaries of the computational domain.

 For the reduced models, a filter has been applied only to the initial condition, and a sponge layer was implemented instead 
 to ensure periodicity of the problem. \\

In all the cases considered in our study, we confirm that the extended cKdV equation performs much better than the leading-order cKdV model. This is especially noticeable in the case of moderate amplitudes, with the amplitude parameters $\eps= 0.3$ and $\eps = 0.5$. Hence, the ecKdV equation extends the range of validity of the weakly-nonlinear modelling to the waves of moderate amplitude, which is the main conclusion of our study. \\

As a by-product of our study, we compared the predictions of Johnson's asymptotic solution for the cKdV equation with the results of our direct numerical simulations and clarified the range of validity of this approximation. Overall, the asymptotics gives good agreement only when  the waves have sufficiently small amplitude (to ensure the applicability of the cKdV model), and the initial condition is placed sufficiently far away from the origin (to ensure that the cylindrical divergence term can be treated as a perturbation of the KdV equation). In addition, the oscillatory transition region approximation is valid only at very large distances away from the initial position of the pulse (indeed, the approximation corresponds to the long-time limit of an undular bore, which depends only on the hight of the shelf), while the approximations to the primary wave and the shelf do not have this limitation.   \\

We also have derived the slow radius versions of the ecKdV equation from the strongly-nonlinear Serre-Green-Naghdi (SGN) and Matsuno models, both for outward and inward propagation, which can be found in the Appendix.\\

We hope that our results will stimulate a renewed interest to the cKdV and extended cKdV models in order to study them from the viewpoint of Hamiltonian structures, near-identity transformations and exact and approximate solutions, similar to those developed for the KdV and extended KdV equations \cite{Ga, ZF, Mag, K, FL}, with subsequent applications to the modelling of concentric waves in various physical contexts. Indeed, these universal models are applicable in all situations with the dominant balance of nonlinearity, dispersion and cylindrical divergence (and similarly, cylindrical convergence for the inward-propagating waves). \\





{\bf Acknowledgements}\\

Nerijus Sidorovas, Dmitri Tseluiko and Karima Khusnutdinova acknowledge the UKRI funding: the PhD project of Nerijus Sidorovas is funded by the Engineering and Physical Sciences Research Council (EPSRC, project reference 2458723). Wooyoung Choi was supported by the US National Science Foundation through Grant No. DMS-2108524. Karima Khusnutdinova and Wooyoung Choi also thank the London Mathematical Society and Loughborough Institute of Advanced Studies for the partial support of their collaborative research.\\


\appendix{\bf Appendix A}\\

In this appendix, we derive the extended cKdV models for stronger nonlinear regimes. Hence, we extend the considerations of our paper to the strongly nonlinear long-wave SGN \cite{Se, GN, SG} and Matsuno's \cite{M1, M2} models for surface waves.
The extended weakly-nonlinear models of the type examined in our paper are more amenable to analysis, and they are likely to provide useful initial conditions for the strongly nonlinear models, similarly to their plane waves counterparts \cite{GBK}.
\\

The dimensionless Matsuno $\delta^4$ model is given by

\begin{equation}
\label{GN2D}
\begin{cases}
\Te_{\Tt} + \tilde{\nabla} \cdot [(1 + \eps \Te) \Tbfu] = 0, \\
\Tbfu_{\Tt} + \eps (\Tbfu \cdot \tilde{\nabla}) \Tbfu + \tilde{\nabla} \Te = \delta^2 S_1 + \delta^4 S_2,
\end{cases}
\end{equation}
where
\begin{align*}
& S_1  = \dfrac{1}{3(1+\eps \Te)} \nabla \left[ (1+\eps \Te)^3 \{ \nabla \cdot \Tbfu_{\Tt} + \eps (\Tbfu \cdot \nabla)(\nabla \cdot \Tbfu) - \eps (\nabla \cdot \Tbfu)^2 \} \right], \\[0.3cm]
\begin{split}
S_2 = \dfrac{1}{45(1 + \eps \Te)}\nabla [ \nabla \cdot \{ (1 + \eps \Te)^5 \nabla(\nabla \cdot \Tbfu_{\Tt}) + \eps (1 + \eps \Te)^5(\nabla^2(\nabla\cdot \Tbfu))\Tbfu -5\eps (1 + \eps \Te)^5 (\nabla \cdot \Tbfu)\nabla (\nabla \cdot \Tbfu) \\ + \eps \nabla h^5 \times (\Tbfu \times \nabla(\nabla \cdot \Tbfu))\}  -3 \eps h^5 \{ \nabla(\nabla \cdot \Tbfu) \}^2 ] \\ - \dfrac{\eps}{45(1 + \eps \Te)} \left[ \nabla \cdot \{ (1 + \eps \Te)^5\nabla(\nabla \cdot \Tbfu) \} \nabla(\nabla \cdot \Tbfu) + \dfrac{(1 + \eps \Te)^5}{2}\nabla \{ \nabla(\nabla \cdot \Tbfu) \}^2 \right].
\end{split}
\end{align*}
Repeating the steps described in Section \ref{sec:cKdV} with the help of symbolic computations in Mathematica \cite{Math} we can derive the ecKdV equation for axisymmetric surface waves. Writing this system is very cumbersome, however (after omitting tildes for brevity) we can convert it to axisymmetric coordinates $(r,t)$ and write it as follows:

\begin{equation}
\label{extendedCKDVsystem}
\begin{cases}
\eta_t + \dfrac{1}{r}[r(1+\eps\eta)U]_r = 0, \\[0.2cm]
U_t + \eps UU_r + \eta_r = \eps \mathcal{S}_1 + \eps^2\mathcal{S}_2,
\end{cases}
\end{equation}
where $S_1,S_2$ transform into the following:
\begin{align}
\label{S1axisymmetric}
\nonumber \mathcal{S}_1 = \dfrac{1 + \eps\eta}{3 r^3} \bigg[ & 3r^3\eps^2\eta_rUU_{rr}-3r^3\eps^2\eta_rU_r^2-r^3\eps^2\eta U_rU_{rr}+r^3\eps^2\eta UU_{rrr}+3r^3\eps\eta_rU_{tr} 
\\ & \nonumber +r^3\eps\eta U_{trr}-r^3\eps U_rU_{rr}+r^3\eps UU_{rrr}+r^3U_{trr}-3r^2\eps^2\eta_rUU_r-r^2\eps^2\eta U_r^2 
\\ & \nonumber -r^2\eps^2\eta UU_{rr}+3r^2\eps\eta_rU_t+r^2\eps\eta U_{tr}-r^2\eps U_r^2-r^2\eps UU_{rr}+r^2U_{tr}
\\ & \nonumber -3r\eps^2\eta UU_r-r\eps\eta U_t-3r\eps UU_r-rU_t-6r\eps^2\eta_rU^2+4\eps^2\eta U^2+4\eps U^2 \bigg],
\end{align}
and
\begin{align}
\nonumber \mathcal{S}_2 = -\dfrac{(1 + \eps\eta)^2}{45 r^5} \bigg[ & -70\eps^2\eta_rU_{rr}^2r^5-70\eps^3\eta\eta_rU_{rr}^2r^5-100\eps^3U_r\eta_r^2U_{rr}r^5-25\eps^2U_rU_{rr}\eta_{rr}r^5
\\ & \nonumber -25\eps^3\eta U_rU_{rr}\eta_{rr}r^5+20\eps^2\eta_r^2U_{trr}r^5+5\eps\eta_{rr}U_{trr}r^5+5\eps^2\eta\eta_{rr}U_{trr}r^5
\\ & \nonumber +20\eps^3U\eta_r^2U_{rrr}r^5-40\eps^2U_r\eta_rU_{rrr}r^5-40\eps^3\eta U_r\eta_rU_{rrr}r^5-22\eps^3\eta^2U_{rr}U_{rrr}r^5
\\ & \nonumber -22\eta U_{rr}U_{rrr}r^5-44\eps^2\eta U_{rr}U_{rrr}r^5+5\eps^2U\eta_{rr}U_{rrr}r^5+5\eps^3U\eta\eta_{rr}U_{rrr}r^5
\\ & \nonumber +10\eps\eta_rU_{trrr}r^5+10\eps^2\eta\eta_rU_{trrr}r^5-3\eps^3\eta^2U_rU_{rrrr}r^5-3\eta U_rU_{rrrr}r^5
\\ & \nonumber -6\eps^2\eta U_rU_{rrrr}r^5+10\eps^2U\eta_rU_{rrrr}r^5+10\eps^3U\eta\eta_rU_{rrrr}r^5+\eps^2\eta^2U_{trrrr}r^5
\\ & \nonumber +2\eps\eta U_{trrrr}r^5+U_{trrrr}r^5+\eps^3U\eta^2U_{rrrrr}r^5+\eta UU_{rrrrr}r^5+2\eps^2U\eta U_{rrrrr}r^5
\\ & \nonumber -100\eps^3U_r^2\eta_r^2r^4-27\eps^3\eta^2U_{rr}^2r^4-27\eta U_{rr}^2r^4-54\eps^2\eta U_{rr}^2r^4+20\eps^2\eta_r^2U_{tr}r^4
\\ & \nonumber -60\eps^3U\eta_r^2U_{rr}r^4-195\eps^2U_r\eta_rU_{rr}r^4-195\eps^3\eta U_r\eta_rU_{rr}r^4-25\eps^2U_r^2\eta_{rr}r^4
\\ & \nonumber -25\eps^3\eta U_r^2\eta_{rr}r^4+5\eta U_{tr}\eta_{rr}r^4+5\eps^2\eta U_{tr}\eta_{rr}r^4-15\eps^2UU_{rr}\eta_{rr}r^4
\\ & \nonumber -15\eps^3U\eta U_{rr}\eta_{rr}r^4+15\eps\eta_rU_{trr}r^4+15\eps^2\eta\eta_rU_{trr}r^4-28\eps^3\eta^2U_rU_{rrr}r^4
\\ & \nonumber -28\eta U_rU_{rrr}r^4-56\eps^2\eta U_rU_{rrr}r^4-25\eps^2U\eta_rU_{rrr}r^4-25\eps^3U\eta\eta_rU_{rrr}r^4
\\ & \nonumber +2\eps^2\eta^2U_{trrr}r^4+4\eps\eta U_{trrr}r^4+2U_{trrr}r^4-2\eps^3U\eta^2U_{rrrr}r^4-2\eta UU_{rrrr}r^4
\\ & \nonumber -4\eps^2U\eta U_{rrrr}r^4-20\eps^2U_{t}\eta_r^2r^3-20\eps^3UU_r\eta_r^2r^3-5\eps^2U_r^2\eta_rr^3-5\eps^3\eta U_r^2\eta_rr^3
\\ & \nonumber -15\eps\eta_rU_{tr}r^3-15\eps^2\eta\eta_rU_{tr}r^3+21\eps^3\eta^2U_rU_{rr}r^3+21\eta U_rU_{rr}r^3+42\eps^2\eta U_rU_{rr}r^3
\\ & \nonumber +45\eps^2U\eta_rU_{rr}r^3+45\eps^3U\eta\eta_rU_{rr}r^3-5\eta U_{t}\eta_{rr}r^3-5\eps^2\eta U_{t}\eta_{rr}r^3-5\eps^2UU_r\eta_{rr}r^3
\\ & \nonumber -5\eps^3U\eta U_r\eta_{rr}r^3-3\eps^2\eta^2U_{trr}r^3-6\eps\eta U_{trr}r^3-3U_{trr}r^3+9\eps^3U\eta^2U_{rrr}r^3
\\ & \nonumber +9\eta UU_{rrr}r^3+18\eps^2U\eta U_{rrr}r^3+30\eps^3\eta^2U_r^2r^2+30\eta U_r^2r^2+60\eps^2\eta U_r^2r^2
\\ & \nonumber +120\eps^3U^2\eta_r^2r^2+15\eta U_{t}\eta_rr^2+15\eps^2\eta U_{t}\eta_rr^2+175\eps^2UU_r\eta_rr^2+175\eps^3U\eta U_r\eta_rr^2
\\ & \nonumber +3\eps^2\eta^2U_{tr}r^2+6\eps\eta U_{tr}r^2+3U_{tr}r^2+9\eps^3U\eta^2U_{rr}r^2+9\eta UU_{rr}r^2+18\eps^2U\eta U_{rr}r^2
\\ & \nonumber +30\eps^2U^2\eta_{rr}r^2+30\eps^3U^2\eta\eta_{rr}r^2-3\eps^2\eta^2U_{t}r-6\eps\eta U_{t}r-3U_{t}r-93\eps^3U\eta^2U_rr
\\ & \nonumber -93\eta UU_rr-186\eps^2U\eta U_rr-170\eps^2U^2\eta_rr-170\eps^3U^2\eta\eta_rr+63\eta U^2+63\eps^3U^2\eta^2
\\ & \nonumber +126\eps^2U^2\eta \bigg].
\end{align}

We now seek solutions in the asymptotic form
\begin{align}
\eta & = \etazero + \eps \etaone + \eps^2 \etatwo + \cdots, \\
U & = \Uzero + \eps \Uone + \eps^2 \Utwo + \cdots,
\end{align}
where all functions $\zeta^{(i)},U^{(i)}$ depend on the fast and slow variables
\begin{equation}
\xi = r - t \quad \text{(fast variable)}, \qquad R = \eps r \quad \text{(slow variable)}.
\end{equation}
We proceed by substituting these expansions into (\ref{extendedCKDVsystem}) and collecting the terms at increasing orders of $\eps$.\\[0.5cm]
At $O(1)$ we obtain the following linearised system for $\etazero,\Uzero$:
\begin{equation}
\begin{cases}
-\etazero_\xi + \Uzero_\xi = 0, \\[0.2cm]
\hphantom{+}\etazero_\xi - \Uzero_\xi  = 0,
\end{cases}
\end{equation}
and from this relation we obtain
\begin{align}
\label{ExtendedCKDVorder1}
\Uzero & = \etazero + \underbrace{f(R)}_{\equiv 0},
\end{align}
where the arbitrary function $f(R)$ must be zero in the framework of rapidly decaying functions such that $\Uzero,\etazero \to 0$ as $\xi \to \infty$ for any fixed $r$. \\[0.5cm]

At $O(\eps)$ we obtain the following system for $\etaone,\Uone$:
\begin{equation}
\label{ExtendedCKDVorder2system}
\begin{cases}
-\etaone_\xi + \Uone_\xi = - \Uzero_{R} - \big[ \etazero\Uzero \big]_{\xi} - \dfrac{1}{R}\Uzero, \\[0.3cm]
 
\hphantom{+}\etaone_\xi - \Uone_\xi = - \etazero_{R} - \dfrac{1}{3}\Uzero_{\xi\xi\xi} - \Uzero \Uzero_{\xi},
\end{cases}
\end{equation}
where we can now take the sum $(\ref{ExtendedCKDVorder2system})_1 + (\ref{ExtendedCKDVorder2system})_2$ and obtain
\begin{equation}
0 = - \Uzero_{R} - [\etazero\Uzero]_{\xi} - \dfrac{1}{R}\Uzero - \etazero_{R} - \dfrac{1}{3}\Uzero_{\xi\xi\xi} - \Uzero \Uzero_{\xi},
\end{equation}
and here we can substitute (\ref{ExtendedCKDVorder1}) to deduce the cKdV equation as
\begin{equation}
\label{ExtendedCKDVorder2}
\etazero_{R} + \dfrac{3}{2} \etazero\etazero_{\xi} + \dfrac{1}{6}\etazero_{\xi\xi\xi} + \dfrac{1}{2R} \etazero = 0.
\end{equation}
The last term describes  the cylindrical divergence. It becomes less significant the further away from the origin the waves travel. We can use the cKdV equation in conjunction with $(\ref{ExtendedCKDVorder2system})_1$ to deduce that

\begin{equation}
\label{U2eta}
\Uone_\xi = \etaone_\xi - \dfrac{1}{2}\etazero\etazero_\xi + \dfrac{1}{6}\etazero_{\xi\xi\xi} - \dfrac{1}{2R}\etazero.
\end{equation}
For the next order, we require to know $\Uone$ and this equation must be integrated through over the region $\xi' \in (\xi,\infty)$ which provides us with the non-local term
\begin{equation}
\label{phione}
\phizero(\xi,R) := -\int_\xi^{\infty} \etazero(\xi',R) \ \D \xi',
\end{equation}
thus putting (\ref{U2eta}) in the form
\begin{equation}
\label{U2}
\Uone = \etaone - \dfrac{1}{4}\etazero\etazero + \dfrac{1}{6}\etazero_{\xi\xi} - \dfrac{1}{2R}\phizero.
\end{equation}
At $O(\eps^2)$ we obtain the following system:

\begin{equation}
\label{ExtendedCKDVorder3system}
\begin{cases}
- \etatwo_{\xi} + \Utwo_{\xi} = - \Uone_{R} - \big[ \etazero\Uzero \big]_R - \big[\etazero\Uone + \etaone\Uzero\big]_{\xi} - \dfrac{1}{R}\etazero\Uzero - \dfrac{1}{R^2}\Uone, \\[0.3cm]
\begin{split}
\etatwo_{\xi} - \Utwo_{\xi} = - \etaone_{R} - \Uzero\Uzero_{R} - \dfrac{2}{3}\Uzero_{R\xi\xi} - \big[\Uzero\Uone\big]_\xi + \dfrac{1}{3}\Uzero\Uzero_{\xi\xi\xi} -\dfrac{2}{3}\etazero\Uzero_{\xi\xi\xi} \\[0.1cm]- \dfrac{1}{45}\Uzero_{\xi\xi\xi\xi\xi} - \dfrac{1}{3}\Uone_{\xi\xi\xi}- \etazero_{\xi}\Uzero_{\xi\xi} - \dfrac{1}{3} \Uzero_{\xi}\Uzero_{\xi\xi} - \dfrac{1}{3R}\Uzero_{\xi\xi}.
\end{split}
\end{cases}
\end{equation}
As before, we take the sum $(\ref{ExtendedCKDVorder3system})_1 + (\ref{ExtendedCKDVorder3system})_2$, and then substitute the following:
\begin{subequations}
\begin{align}
& \text{From (\ref{ExtendedCKDVorder1})}: & \Uzero & = \etazero, \label{sub1} \\[0.3cm]
&\text{From (\ref{U2})}: & \Uone & = \etaone - \dfrac{1}{4}[\etazero]^2 + \dfrac{1}{6}\etazero_{\xi\xi} - \dfrac{1}{2R}\phizero, \label{sub2} \\[0.3cm]
&\text{From (\ref{ExtendedCKDVorder2})}: & \etazero_R & = -\dfrac{3}{2}\etazero\etazero_{\xi} - \dfrac{1}{6}\etazero_{\xi\xi\xi} - \dfrac{1}{2R}\etazero, \label{sub3} \\[0.3cm]
&\text{From (\ref{ExtendedCKDVorder2})}: & \etazero_{R\xi\xi} & = -\dfrac{9}{2} \etazero_{\xi} \etazero_{\xi\xi} -  \dfrac{3}{2} \etazero\etazero_{\xi\xi\xi} - \dfrac{1}{6} \etazero_{\xi\xi\xi\xi\xi} - \dfrac{1}{2R} \etazero_{\xi\xi}, \label{sub4} \\[0.3cm]
&\text{From (\ref{phione})}: & \phizero_\xi & = \etazero, \label{sub5} \\[0.3cm]
&\text{From (\ref{phione})}: & \phizero_R & = -\int_{\xi}^{\infty} \etazero_R (\xi',R) \ d\xi' = -\dfrac{3}{4}\big[ \etazero \big]^2 - \dfrac{1}{6}\etazero_{\xi\xi} - \dfrac{1}{2R}\phizero. \label{sub6}
\end{align}
\end{subequations}
This significantly reduces (\ref{ExtendedCKDVorder3system}) to
\begin{align}
\label{EcKdVphiRphi}
\etaone_R + \dfrac{3}{2}\bigg[ \etazero\etaone\bigg]_\xi + \dfrac{1}{6}\etaone_{\xi\xi\xi} + \dfrac{1}{2R}\etaone - \bigg( \dfrac{21}{8}&\big[\etazero\big]^2\etazero_{\xi} + \dfrac{31}{24}\etazero_{\xi}\etazero_{\xi\xi} + \dfrac{7}{12}\etazero\etazero_{\xi\xi\xi} + \dfrac{11}{360}\etazero_{\xi\xi\xi\xi\xi} \nonumber \\[0.1cm] & + \dfrac{1}{16R}\left[ 9 \big(\etazero\big)^2 + 8\phizero \etazero_\xi \right] - \dfrac{1}{8R^2}\phizero \bigg) = 0.
\end{align}
We proceed to multiply equation (\ref{EcKdVphiRphi}) by $\eps$ and add on equation (\ref{ExtendedCKDVorder2}). It prompts us to consider $\He = \etazero + \eps \etaone$ and omit any terms of order $\eps^2$ or higher. Doing so yields the ecKdV model in the form
\begin{align}
\label{extendedCKDVrelationMATSUNO}
\He_R + \dfrac{3}{2} \He\He_\xi + \dfrac{1}{6}\He_{\xi\xi\xi} + \dfrac{1}{2R}\He &-\eps \bigg( \dfrac{21}{8}\He^2 \He_\xi + \dfrac{7}{12}\He\He_{\xi\xi\xi} + \dfrac{31}{24}\He_\xi \He_{\xi\xi} + \dfrac{11}{360} \He_{\xi\xi\xi\xi\xi} +\dfrac{1}{16R}\bigg[ 9\He^2 + 8\He_\xi \Hphi \bigg] - \dfrac{1}{8R^2}\Hphi \bigg) = 0.
\end{align}
By omitting all $O(\delta^4)$ terms from \eqref{GN2D}, we obtain the SGN equations and the ecKdV is derived in the form 

\begin{equation}
\label{GNeckdv}
\He_R + \dfrac{3}{2} \He\He_\xi + \dfrac{1}{6}\He_{\xi\xi\xi} + \dfrac{1}{2R}\He -\eps \bigg( \dfrac{21}{8}\He^2 \He_\xi + \dfrac{7}{12}\He\He_{\xi\xi\xi} + \dfrac{31}{24}\He_\xi \He_{\xi\xi} + \dfrac{1}{24}\He_{\xi\xi\xi\xi\xi} +\dfrac{1}{16R}\bigg[ 9\He^2 + 8\He_\xi \Hphi \bigg] - \dfrac{1}{8R^2}\Hphi \bigg) = 0.
\end{equation} 

 
 We can take the reduction even further by omitting $O(\eps \delta^2)$ terms from \eqref{GN2D} instead to obtain the Boussinesq system \eqref{Boussinesq2Ddimless}, and the corresponding ecKdV equation was derived in Section \ref{sec:cKdV}.\\


We also note that the inward-propagating ecKdV models for the axisymmetric versions of 2D Boussinesq, SGN, and Matsuno systems require introduction of another characteristic variable $\bar{\xi} = \tilde{r}+\tilde{t}$ and give rise to the equation
\begin{equation}
\He_R - \dfrac{3}{2} \He\He_{\bar{\xi}} - \dfrac{1}{6}\He_{\bar{\xi}\bar{\xi}\bar{\xi}} + \dfrac{1}{2R}\He + \eps \bigg( \dfrac{21}{8}\He^2 \He_{\bar{\xi}} + A_1\He_{\bar{\xi}} \He_{\bar{\xi}\bar{\xi}} + A_2\He\He_{\bar{\xi}\bar{\xi}\bar{\xi}} + A_3\He_{\bar{\xi}\bar{\xi}\bar{\xi}\bar{\xi}\bar{\xi}} - \dfrac{1}{16R}\bigg[ 9\He^2 + 8\He_{\bar{\xi}} \Hphi \bigg] + \dfrac{1}{8R^2}\Hphi \bigg) = 0.
\end{equation}
The constants $(A_1,A_2,A_3)$ take on the values of $(47/24,3/4,1/24)$ in the case of Boussinesq system, \linebreak  $(31/24,7/12,1/24)$ for the SGN system, and $(31/24,7/12,11/360)$ for the Matsuno system.




\end{document}